\documentclass[final]{siamltex}

\usepackage[colorlinks=true,urlcolor=blue]{hyperref}
\usepackage{amstext}
\usepackage{amssymb}
\usepackage{amsmath}
\usepackage{mathrsfs}
\usepackage{array}
\usepackage{enumerate}
\usepackage{url}
\usepackage{mathtools}

\usepackage[]{graphicx}               
\usepackage{epstopdf}
\usepackage{color}
\usepackage{appendix}
\usepackage[section]{placeins}
\usepackage[section]{algorithm}
\usepackage{algorithmicx}
\usepackage{algpseudocode}

\usepackage{tikz}
\usetikzlibrary{shapes,arrows}

\tikzstyle{decision} = [diamond, draw, text width=4.5em, text badly centered, node distance=3cm, inner sep=0pt]
\tikzstyle{block} = [rectangle, draw, text width=8em, text centered, rounded corners, minimum height=4em]

\newcommand{\reals}{\mathbb{R}}
\newcommand{\vecx}{\vec{x}}
\newcommand{\DD}{\displaystyle}
\DeclareMathOperator{\sech}{sech}
\def\multiset#1#2{\ensuremath{\left(\kern-.3em\left(\genfrac{}{}{0pt}{}{#1}{#2}\right)\kern-.3em\right)}}

\usepackage{verbatim}  

\title{Curvature-driven molecular flows on membrane surfaces 
\thanks{This work has been partially supported by the Simons Foundation and National Institutes of 
Health through the grant R01GM11759301 as part of the joint DMS/NIGMS initiative to support research at the interface of the biological 
and mathematical sciences.}}

\author{Michael Mikucki\thanks{Department of Mathematics,
        Colorado School of Mines, Golden, Colorado, 80401-1887
        ({\tt mikucki@mines.edu}).}
        \and Y.~C.~Zhou\thanks{Department of Mathematics,
        Colorado State University, Fort Collins, Colorado, 80523-1874
        ({\tt yzhou@math.colostate.edu}).}}

\begin{document}

\maketitle

\begin{abstract}
Morphological change of bilayer membrane {\it in vivo} is not a spontaneous procedure but modulated by various
types of proteins in general. Most of these modulations are associated with the localization of related proteins in the 
crowded lipid environment in bilayer membrane. This work presents an mathematical model for the localization of multiple 
species of diffusion molecules on membrane surfaces. We start with the energetic description of the distributions of 
molecules on curved membrane surface, by assembling the bending energy of bilayer membrane and the entropic energy of
diffusive molecules. We introduce the spontaneous curvature of molecules in membrane, and define the spontaneous curvature
of bilayer membrane as a function of the molecule concentrations on membrane surfaces. This connection gives rise to 
a drift-diffusion equation to govern the gradient flows of the surface molecule concentrations. We recast the energetic
formulation and the related governing equations in the Eulerian framework by using a phase field function that defines
the membrane morphology. Computational simulations with the proposed mathematical model and related numerical 
techniques predict the molecular localization on membrane surfaces at locations with preferred mean
curvature. 
\end{abstract}



\pagestyle{myheadings}
\thispagestyle{plain}
\markboth{MICHAEL MIKUCKI AND Y. C. ZHOU}{CURVATURE-DRIVEN PROTEIN FLOW}

\section{Introduction}

This paper concerns the derivation of a curvature-driven diffusion equation for membrane bound proteins.  Experiments suggest that diffusive proteins within 
lipid membranes play a significant role in producing and regulating membrane curvature 
\cite{Baumgart2011, Zimmerberg2006, Antonny2011, Cui2011, Parton2011, Stachowiak2010, Huttner2001}.  We 
capture a major class of these effects in our model, which couples the lateral protein diffusion with 
the dynamic membrane along which these proteins are located.

Approximately 30-90\% of all membrane proteins can freely diffuse along the membrane \cite{Faraudo2002, Kim1998}.  In addition, proteins 
induce various curvatures to the underlying membrane.  A few mechanisms of protein induced curvature are summarized as follows.  
Rigid proteins such as those in the BAR (Bin/Amphiphysin/Rvs) domain family can act as a scaffold to the membrane.  
These proteins have an intrinsic curvature and, upon attachment, the membrane bends to match the protein curvature.  In 
a similar fashion, several proteins can oligomerize to create a rigid shape and bend the membrane.  Protein 
coats such as clathrin, COPI ({\sc CO}at {\sc P}rotein {\sc I}) and COPII ({\sc CO}at {\sc P}rotein {\sc II}) are examples of this type.  
Another type of protein induced membrane curvature is through protein insertion.  Membrane 
curvature is induced when there is a difference between the length of the hydrophobic region of a membrane 
protein and the thickness of the hydrophobic core of the lipid bilayer in which it is embedded \cite{Parton2011}.  Epsin 
proteins do this by forming an $\alpha$-helix upon binding to the membrane, and this helix, known as H0, 
inserts itself into the membrane \cite{Baumgart2011}.  Moreover, local protein crowding of peripheral proteins can cause 
membrane bending by creating an asymmetry of the monolayer areas and thereby curling the membrane away from the side 
which the crowding occurred.  This effect is experimentally demonstrated in \cite{Stachowiak2010}.  Further illustrating the importance of 
proteins in membranes, Schmidt {\it et. al.} showed that the M2 protein plays an essential role in generating 
regions of high curvature in the influenza A virus membrane \cite{Schmidt2013}.  This specific protein accumulates 
in regions of negative Gaussian curvature and can generate curvature in the membrane itself, creating a positive feedback loop 
and allowing the virus to replicate.  
We finally note that all endocytosis and exocytosis processes are promoted in one way or another by proteins.  Therefore, 
any viral replication process requires proteins.  Antagonizing these curvature effects of proteins may be a viable 
antiviral strategy \cite{Schmidt2013}.  This motivates the necessity for a model coupling membrane shape and protein diffusion.  

The classical mechanical bending energy of a bilayer membrane, given by Canham \cite{Canham1970}, Helfrich, \cite{Helfrich1973}, and Evans \cite{Evans1974} 
depends only upon the membrane curvature.  However, when a force that produces a topological change to the membrane surface, such as those 
induced by proteins, this so-called sharp-interface model fails, since a change to the topology creates a discontinuity in the energy 
functional.  In addition, a change in topology requires a discontinuous surface for a moment, which is impossible to 
model using an explicit parameterization of the surface.  An effective way to treat topological changes is to track 
the surface implicitly as a level set of a 3D function.  This method is called the phase field method or 
diffuse-interface method and has been very successful in modeling membrane dynamics \cite{Du2004, Du2006, Teigen2011}.  The membrane is defined 
by a level set of a phase field function, $\phi$, and the motion of the membrane is governed by gradient flow of the energy 
functional, ensuring a decrease in energy in time.  Since the membrane is never explicitly tracked, topological changes 
can occur.  This approach has the obvious advantage over sharp interface models since it provides a way 
to describe topological changes to the membrane.  However, a disadvantage of this method is that it is difficult 
for this method to describe local forces on the membrane.  If local forces need to be modeled, one should 
use the sharp-interface method.  Furthermore, this method requires solving a higher dimensional system, since the energy functional is 
computed over the entire space $\Omega \subset \reals^3$ rather than just a manifold $\Gamma\subset \reals^2$.  This is 
the price we must pay to describe membrane merging and separating.

We need to use a phase field approach to model the dynamics of the membrane surface when membranes merge and separate.  
We must also include the activity of proteins in our phase field model.  Du {\it et. al.} have successfully used a phase field approach to 
track multiple diffusive lipid species by using two phase functions \cite{Du2008}. 
The phase functions are orthogonal and their intersections define the separation of the two lipid species.  They have been able to 
reproduce numerous vesicle shapes which match experimental results \cite{Baumgart2003}.  However, we argue that a phase field 
approach should not be used to track the dynamics of the diffusive membrane proteins.  Lipid species may arrange 
themselves into distinct phases, but proteins do not necessarily form separate phases \cite{Kim1998}.  Therefore, a 
dual phase field model cannot account for the effect of diffusive proteins in lipid membranes.  We describe the 
proteins as diffusive particles governed by the advection-diffusion equation.  A continuum model for the 
diffusion of proteins is physically justifiable by the relative length scales of the proteins embedded 
in the membrane, which are typically 4-5nm.\ thick, to the cell, which can be up to 100 $\mu$m in diameter.  
(In the aforementioned virus replication example, the spherical virions produced from the budding are typically 100nm in diameter \cite{Rossman2010}.)  
At such length scales, we may consider the proteins as diffusive particles that are attracted toward regions of 
specific membrane curvature.  Therefore, the membrane proteins follow an equation similar to 
the usual drift-diffusion equation.  The difference between this diffusion equation and the usual drift-diffusion equation is 
that the flux is proportional to a diffusion potential which is governed by the curvature energy of the membrane.  
That is, it is the curvature of the membrane responsible for the ``drift'' of the proteins.    

The diffusion of the proteins occurs only on the membrane surface, which is implicitly defined in a phase-field framework.  Various 
techniques have been established for solving PDEs on surfaces, and we provide a brief overview here.  Dziuk used finite 
element methods to solve elliptic partial differential equations on stationary surfaces \cite{Dziuk1988}.  This work was expanded with Elliott for parabolic 
equations on dynamic surfaces \cite{Dziuk2010}.  This work was extended even further with Deckelnick and Heine by solving the 
equation using a narrow band around the surface \cite{Deckelnick2010}.  Other approaches for PDEs on surfaces 
include \cite{Teigen2011, Greer2006, Adalsteinsson2003, Zhao2003, Olshanskii2015} and references therein.  In all of these approaches, 
mesh refinement is required to accurately resolve the numerical solution.  Therefore, these methods are suitable for 
stationary surfaces, but mesh refinement for dynamic surfaces can be a significant computational challenge.  

For this reason, we use the Fourier spectral method to solve the diffusion equation.  This method is also 
available to solve the phase field equations \cite{Du2004, Du2006, Du2008, Teigen2011}, giving a consistent solution procedure.  Fourier 
methods exhibit exponential convergence, meaning that the error decreases faster than any power of the grid size \cite{Strain1994}.  Therefore, 
local mesh refinement near the interface is less important in Fourier (global) approaches than finite element and 
finite difference (local) approaches.  With this consistent framework, we may simultaneously solve for the 
shape of the membrane and the dynamics of the diffusive proteins in the membrane.  The results of this coupled procedure 
produce effects that are not easily observable in experiments.

This paper is organized as follows.  First, the total energy for the system is defined in a phase field framework 
in Section \ref{sec:membrane}.  A curvature-driven diffusion equation is defined for the diffusive proteins 
in the same framework in Section \ref{sec:protein}.   The presence of the proteins induce a curvature on the 
membrane through the spatially variable and concentration dependent spontaneous curvature.  A solution procedure 
for the resulting PDEs using Fourier spectral methods will be presented in a future article.  
Finally, results for diffusive proteins exhibiting various curvature preferences on the surface of a torus are presented in Section \ref{sec:results}.


\section{Energy formulation}\label{sec:membrane}
We model an enclosed bilayer membrane as a structure-less surface $\Gamma$ contained in a three-dimensional domain $\Omega \in \mathbb{R}^3$. 
The membrane $\Gamma$ separates $\Omega$ into two subdomains, one inside the membrane and the other outside. On the membrane there distribute
$m+1$ distinct lipid species with concentrations $\rho^{\rm lip}_l$, $0 = 1,\dots, m$, and a single 
diffusive membrane protein with concentration $\rho^{\rm pro}$. Throughout our notation, we use subscripts to denote 
the species number and superscripts to denote the species type. We consider only one protein species in this work, while 
the model can easily be extended with the use of subscripts. The total energy of the system is composed of the membrane bending 
energy in Eulerian form, which includes the effects of the multiple lipid and protein species, and the entropic energy from the 
lipids and proteins: 
\begin{equation}\label{eq:E-tot}
E_{\rm tot} = E_{\rm mem} + E_{\rm ent}.
\end{equation}
The exact representation of the total energy depends on the representation of the membrane $\Gamma$, which can
be given explicitly as a parameterized three-dimensional surface or implicitly as a level set of a three-dimensional function that 
is defined in the entire domain $\Omega$. The corresponding energy formulations using the explicitly and implicitly represented
surfaces will be referred to as the Lagrangian and Eulerian formulations, respectively. The Lagrangian formulation is a direct
mathematical description of the energetic nature of the interacted protein-membrane system. However, numerical implementations 
of this formulation in tracing the dynamics of the membrane suffers from the geometrical singularities that may arise when
there is a topological change in membrane morphology. By using the Eulerian formulation one can track the dynamics of the
membrane $\Gamma$ by evolving the underlying three-dimensional function in the entire domain $\Omega$. In this work we
first introduce the total energy using the Lagrangian formulation and then translate it into the Eulerian formulation
for ease of exposition. 
\subsection{Lagrangian Formulation}
The Lagrangian form of the membrane energy $E_L$ is the Canham-Helfrich-Evans membrane energy derived from 
fundamental physical principles \cite{Helfrich1973, Frank1958}:
\begin{equation}\label{eq:EL}
E_{\rm mem} = \DD\int_\Gamma k(H-C_0)^2 ds,
\end{equation}
where $H$ is the mean curvature of the membrane $\Gamma$, $C_0$ is the spontaneous curvature of the membrane, and $k$ is the bending modulus.  
We note that the above equation neglects surface tension and stretching rigidity.  The surface tension is constant in vesicles with fixed 
surface area giving justification of our simplification \cite{Du2006}.  We refer the reader to \cite{Du2004} for adding stretching rigidity 
to \eqref{eq:EL}. The spontaneous curvature is an intrinsic property of the lipid composition of membrane \cite{Farsad2003}, and when proteins
are induced in the bilayer, it should depend on the protein structure and distribution as well \cite{SodtA2014a,Baumgart2011}. We are motivated by this
biophysical nature to model the membrane spontaneous as a local parameter that depends on the surface density of lipids and proteins. 
Each lipid species $l$ has an intrinsic spontaneous curvature associated to it, denoted $C_0^{l}$.  Furthermore, proteins can induce membrane curvature 
when embedded in a membrane, and a corresponding spontaneous curvature of the protein $C_0^{\rm pro}$ can be 
measured \cite{Callenberg2012}. We define $C_0(\rho_l^{\rm lip},\rho^{\rm pro})$ as the average of the spontaneous curvatures of the contributing species
weighted by their respective fractions of surface coverage:
\begin{equation}\label{eq:C0-all1} 
C_0 = \sqrt{2}\left(\frac{\DD\sum_{l=0}^m C_0^l (a_l^{\rm lip})^2 \rho_l^{\rm lip} + C_0^{\rm pro} (a^{\rm pro})^2 \rho^{\rm pro}}{\DD\sum_{l=0}^m (a_l^{\rm lip})^2 \rho_l^{\rm lip} + (a^{\rm pro})^2 \rho^{\rm pro}} \right),
\end{equation}
where the spontaneous curvatures $C_0^l$ and $C_0^{\rm pro}$ are constants pertaining to the lipid and protein structures. 
The $a_l^{\rm lip}$ are the effective sizes of lipids for $l=1,\dots, m$.  Each lipid is modeled as a hard disk occupying some surface area in the 
membrane, hence we take $(a_l^{\rm lip})^2$ for 
an effective surface area.  Similarly, the $a^{\rm pro}$ is the effective size of the protein embedded in the membrane, occupying 
some surface area $(a^{\rm pro})^2$.  The concentration of particles on the membrane cannot exceed the available space, 
so the concentrations must satisfy the saturation condition
\begin{equation}\label{eq:lipid-cons}
\sum_{l = 0}^m (a_l^{\rm lip})^2 \rho_l^{\rm lip} + (a^{\rm pro})^2 \rho^{\rm pro}  = 1. 
\end{equation}
With this condition \eqref{eq:lipid-cons} the spontaneous curvature defined by \eqref{eq:C0-all1} can be simplified to be
\begin{equation}
C_0 = \sqrt{2} \left( \DD\sum_{l=0}^m C_0^l (a_l^{\rm lip})^2 \rho_l^{\rm lip} + C_0^{\rm pro} (a^{\rm pro})^2 \rho^{\rm pro} \right).
\end{equation}

The entropic energy for the membrane with embedded proteins is defined following the Boltzmann relation by
\begin{equation} \label{eq:E-ent}
E_{\rm ent} = \frac{1}{\beta} \int_{\Gamma} \left(\sum_{l=0}^m \rho_l^{\rm lip} \left[\ln\left(\rho_l^{\rm lip} (a_l^{\rm lip})^2\right)-1\right]
+ \rho^{\rm pro} \left[\ln\left(\rho^{\rm pro} (a^{\rm pro})^2\right)-1\right] \right )  ds, 
\end{equation}
where $\beta = 1/(k_BT)$ is the inverse thermal energy \cite{Zhou2012}.  

The variation of the surface concentration of lipids and proteins follows the general mass conservation law. It reads for a general concentration $\rho$ 
on $\Gamma$ that
\begin{equation}\label{eq:mass-cons}
\frac{\partial \rho}{\partial t} + (\nabla_{s} \cdot \mathrm{v})  \rho = - \nabla_s \cdot \mathrm{J},
\end{equation}
if the surface $\Gamma$ evolves with a normal velocity of $\mathrm{v} \cdot \mathrm{n}$, where $\mathrm{v}$ is a divergence-free velocity field in $\Omega$, 
$\nabla_s \cdot$ is the surface divergence, and $\mathrm{J}$ is the flux vector on surface. Divergence-free velocity field is what the membrane experiences since 
the fluid in which it is immersed is incompressible. We note that the surface advection-diffusion equation of the form
\begin{equation*} 
\frac{\partial \rho}{\partial t} + {\rm v} \cdot \nabla_s \rho = - \nabla_s \cdot \mathrm{J} 
\end{equation*}
is less relevant to the transportation of lipids or proteins on membrane surface, because it assumes a steady surface in the velocity 
field ${\rm v}$ thereby $\mathrm{v} \cdot \mathrm{n} =0$ and $\nabla \cdot \mathrm{v} = \nabla_s \cdot \mathrm{v}$.  
A constitutive relation for the flux is given by the Nernst-Planck formula as an extension 
of Fick's first law,
\begin{equation}\label{eq:flux}
\mathrm{J} = - D_{\Gamma} \beta \rho \nabla_s \mu,
\end{equation}
where $\mu$ is the diffusion potential, $D_{\Gamma}$ is the (constant) lateral diffusion coefficient \cite{AlmeidaP1995a,AlmeidaP1995a}, and $\nabla_s$ 
is the surface gradient \cite{Zhou2012}.  
The diffusion potential $\mu$ is defined as the variation of the total energy with respect to the corresponding surface concentration: 
\begin{align}
\mu^{\rm lip}_l &= \DD\frac{\delta E_{\rm tot}}{\delta \rho_l^{\rm lip}} = \DD\frac{\delta E_{\rm mem}}{\delta \rho_l^{\rm lip}} + \DD\frac{\delta E_{\rm ent}}{\delta \rho_l^{\rm lip}} ,   \qquad l = 1, \dots, m ;\label{eq:mu-lip1}\\
\mu^{\rm pro} &= \DD\frac{\delta E_{\rm tot}}{\delta \rho^{\rm pro}} = \DD\frac{\delta E_{\rm mem}}{\delta \rho^{\rm pro}}  + \DD\frac{\delta E_{\rm ent}}{\delta \rho^{\rm pro}}.  \label{eq:mu-pro1}
\end{align}
It is not necessary to solve a PDE for the concentration of the $0^{th}$ species of lipids thanks to the saturation condition \eqref{eq:lipid-cons}. 
For the computation of the entropic portion of the diffusion potentials, we solve \eqref{eq:lipid-cons} for $\rho_0 ^{\rm lip}(a_0^{\rm lip})^2$ and 
substitute in the entropic energy \eqref{eq:E-ent} to obtain an easier form for differentiating. We shall have
\begin{align}
E_{\rm ent} &= \frac{1}{\beta} \int_{\Gamma} \Bigg( \frac{1}{\big(a_0^{\rm lip}\big)^2}\left(1-\rho^{\rm pro}(a^{\rm pro})^2 - \sum_{l=1}^m \rho_l^{\rm lip}(a_l^{\rm lip})^2  \right) \notag \\
           &  \qquad \times \left[ \ln\left( 1-\rho^{\rm pro}(a^{\rm pro})^2 - \sum_{l=1}^m \rho_l^{\rm lip}(a_l^{\rm lip})^2 \right) -1  \right] +  \notag \\
           & \qquad  +  \sum_{l=1}^m \rho_l^{\rm lip} \left[\ln\left(\rho_l^{\rm lip} (a_l^{\rm lip})^2\right)-1\right] + \rho^{\rm pro} 
\left[\ln\left(\rho^{\rm pro} (a^{\rm pro})^2\right)-1\right]\Bigg) ~ ds. \label{eq:E-ent-lip}
\end{align}
For the lipid species, the derivative is, for each $l = 1, \dots, m$,
\begin{align}
&\frac{\delta E_{\rm ent}}{\delta \rho_l^{\rm lip}}  = \frac{1}{\beta} \Bigg( \frac{1}{(a_0^{\rm lip})^2} \left( -(a_l^{\rm lip})^2 \right) 
\left[ \ln\left(1-\rho^{\rm pro}(a^{\rm pro})^2-\sum_{j=1}^m \rho_j^{\rm lip} (a_j^{\rm lip})^2\right) - 1 \right] \notag\\
    & \quad + \frac{1}{(a_0^{\rm lip})^2}\left( 1-\rho^{\rm pro}(a^{\rm pro})^2 -\sum_{j=1}^m \rho_j^{\rm lip} (a_j^{\rm lip})^2)  \right) 
\left[ \frac{-(a_l^{\rm lip})^2}{1- \rho^{\rm pro}(a^{\rm pro})^2 - \displaystyle{ \sum_{j=1}^m} \rho_j^{\rm lip} (a_j^{\rm lip})^2}  \right]  \notag \\
    & \quad + \left[ \ln(\rho_l^{\rm lip} (a_l^{\rm lip})^2) - 1\right] + \rho_l^{\rm lip} \frac{(a_l^{\rm lip})^2}{\rho_l^{\rm lip} (a_l^{\rm lip})^2} \Bigg) \notag \\
    & = \frac{1}{\beta} \Bigg( \frac{-(a_l^{\rm lip})^2}{(a_0^{\rm lip})^2}  \ln\left(1-\rho^{\rm pro}(a^{\rm pro})^2-\sum_{j=1}^m \rho_j^{\rm lip} (a_j^{\rm lip})^2\right) + \ln(\rho_l^{\rm lip} (a_l^{\rm lip})^2) \Bigg). \label{eq:entropicpart} 
\end{align}
The entropic portion of the diffusion potential for the protein species is computed similarly.

The curvature-driven portion of the diffusion potential is defined as the variation of the membrane energy with respect to the concentrations. 
Using the Lagrangian formulation given in Eq.\eqref{eq:EL}, we compute the variation to be  
\begin{align} 
\frac{\delta E_{\rm mem}}{\delta \rho_l^{\rm lip}} & = 2 k (C_0 - H) \frac{\partial C_0}{\partial \rho_l^{\rm lip}} = 2 k C_0^l (a_l^{\rm lip})^2(C_0 - H),  
\label{eq:dELdp_lip} \\
\frac{\delta E_{\rm mem}}{\delta \rho^{\rm pro}} & = 2 k (C_0 - H) \frac{\partial C_0}{\partial \rho^{\rm pro}} = 2 k C_0^l (a_l^{\rm lip})^2(C_0 - H).  
\label{eq:dELdp_pro} 
\end{align}
The full potential for the individual species of lipids is then given by combining Eqs.(\ref{eq:entropicpart}-\ref{eq:dELdp_lip}),

\begin{align}
\mu_l^{\rm lip} &= \frac{1}{\beta} \left[ \frac{-(a_l^{\rm lip})^2}{(a_0^{\rm lip})^2}  
\ln\left(1-\rho^{\rm pro}(a^{\rm pro})^2-\sum_{j=1}^m \rho_j^{\rm lip} (a_j^{\rm lip})^2\right) + \ln(\rho_l^{\rm lip} (a_l^{\rm lip})^2) \right] + \notag \\
 & \quad 2C_0^l (a_l^{\rm lip})^2 (C_0 - H) \label{eq:mu-lip-Lag}
\end{align}
and the transportation equation can be obtained by using this full diffusion potential to the prototype equation \eqref{eq:mass-cons}. The transportation 
equation for proteins can be obtained similarly. 
\subsection{Eulerian Formulation}
\begin{figure}[!ht]
    \centering
   \includegraphics[height=4cm]{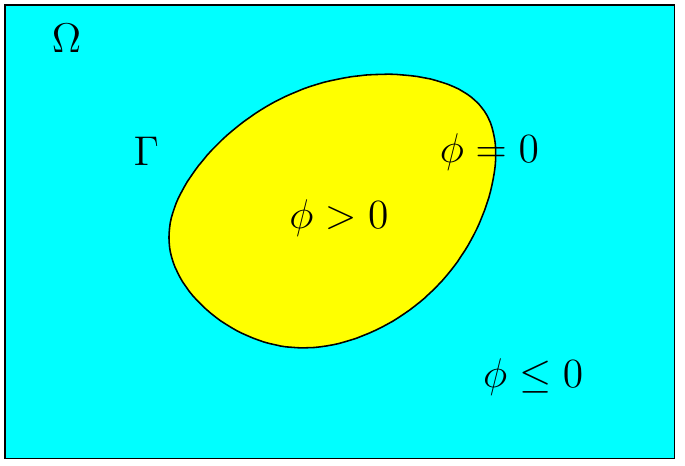} \hspace{5mm}
   \caption{Properties of phase function $\phi$ within the domain $\Omega$.} 
   \label{fig:domain} 
\end{figure}

We anticipate that the molecular localization model being developed here can be finally coupled to dynamic morphological change
of membrane where the position of membrane surface $\Gamma$ is unknown {\it a priori}. For that purpose it is desirable to replace
the above Lagrangian formulation with an alternative formulation that is independent of the parameterization of the surface $\Gamma$.
Here we recast the energy functional in an Eulerian formulation through a smooth phase field function $\phi$. The zero level set of $\phi$ 
separates $\Omega$ into two subdomains, $\Omega_i$ in the interior of $\Gamma$ and $\Omega_e$ to the exterior of $\Gamma$. 
In other words, the set $\{x : \phi(x) = 0 \}$ represents the membrane $\Gamma$, the set $\{ x : \phi(x) > 0\}$ 
represents points inside the membrane, $x\in \Omega_i$, and the set $\{x : \phi(x) \leq 0\}$ represents points outside 
the membrane, $x\in \Omega_e$.  These properties are illustrated in Figure \ref{fig:domain}.  

We use the phase field function as that in \cite{Du2006},
\begin{equation}\label{eq:phi}
\phi(x) = \tanh\left(\frac{d(x)}{\sqrt{2}\epsilon}\right).
\end{equation}
The function $d(x)$ is a signed distance for any point $x \in \Omega$ to the surface $\Gamma$.  It satisfies the 
properties that $d(x) = 0$ for $x \in \Gamma$, $d(x) > 0$ for $x \in \Omega_i$ and $d(x) < 0$ for $x \in \Omega_e$.  
The parameter $0<\epsilon \ll 1$ adjusts the transition width of $\phi$ across the membrane.  
The dilation $1/(\sqrt{2}\epsilon)$ for small $\epsilon$ gives $\phi$ a steep sigmoid shape.  The 
sharp-interface is recovered in the limit as $\epsilon \to 0$.  Additional reasons for the choice of $\tanh(\cdot)$ are explained in \cite{Du2006}.  

The equivalent Eulerian form of the membrane energy $E_{\rm mem}$ is given by
\begin{equation}\label{eq:E-mem}
E_{\rm mem} = \DD\int_\Omega \frac{\epsilon}{2} \left| \Delta \phi - \frac{1}{\epsilon^2} (\phi^2 - 1)(\phi + 
C_0(\rho_l^{\rm lip}, \rho^{\rm pro}) \epsilon) \right|^2  dx,
\end{equation}
where $C_0$ is the spontaneous curvature of the membrane as defined in \eqref{eq:C0-all1}, accounting for the dependence of membrane curvature energy on
the local concentrations of lipids and proteins. If we substitute \eqref{eq:phi} into \eqref{eq:E-mem} with the above-defined phase field function and 
take the limit $\epsilon \to 0$, the Lagrangian formulation \eqref{eq:EL} will be recovered \cite{Du2005}. The scaling by $\sqrt{2}$ appears from this 
derivation of the equivalence to the two-dimensional energy in the sharp interface limit.  

In addition to the Eulerian formulation for the bending energy, we define the following area and volume constraints. The volume constraint is given by 
\begin{equation}\label{eq:vol_constr}
A(\phi) = \int_\Omega \phi(x) \; dx.
\end{equation}
Notice that the integral defined by \eqref{eq:vol_constr} gives
\begin{equation*}
\int_\Omega \phi \; dx = \int_{\Omega_i} \phi \; dx + \int_{\Omega_e} \phi \; dx \overset{\mathclap{\strut{\epsilon\to0}}}\longrightarrow \int_{\Omega_i} 1 \; dx + \int_{\Omega_e} -1 \; dx.
\end{equation*}
That is, $A(\phi)$ approaches the difference between the interior and exterior volumes. The surface area constraint is defined by
\begin{equation}\label{eq:surf_constr}
B(\phi) = \int_\Omega \left(\frac{\epsilon}{2} \left| \nabla \phi \right|^2 + \frac{1}{4\epsilon} (\phi^2 - 1)^2 \right)\; dx.
\end{equation}
For small $\epsilon$, the integrand of \eqref{eq:surf_constr} is significant only near the interface $\Gamma$, and as $\epsilon \to 0$, 
we have $B(\phi) \to 2\sqrt{2}/3\cdot {\rm area}(\Gamma)$ \cite{Du2005}.


\section{Curvature-driven Transportation}  \label{sec:protein}

We proceed to the treatment of the diffusive lipids and proteins with the new Eulerian formulation of total energy. 
While the diffusion occurs on the membrane where $\phi=0$, solving diffusion equations on the surface involves
surface mesh generation, mesh deformation, and re-meshing if there is a large deformation or topological change 
to the membrane surface during the evolution of the phase field function $\phi$. We choose to extend the diffusion
domain from $\Gamma$ to the entire $\Omega$, practically concentrated in the neighborhood of $\Gamma$ following
the evolution of $\phi$, to avoid the treatment of a surface mesh. Correspondingly, the Lagrangian formulation of
the entropic energy in \eqref{eq:E-ent} need to the replaced by the following Eulerian formulation
\begin{equation} \label{eq:E-ent-Eulerian}
E_{\rm ent} = \frac{1}{\beta} \int_{\Omega} \left(\sum_{l=0}^m \rho_l^{\rm lip} \left[\ln\left(\rho_l^{\rm lip} (a_l^{\rm lip})^2\right)-1\right]
+ \rho^{\rm pro} \left[\ln\left(\rho^{\rm pro} (a^{\rm pro})^2\right)-1\right] \right ) dx. 
\end{equation}
This formulation indicates that the surface distributions of lipids and proteins need to be replaced by 
volume distributions. The initial volume distribution can be obtained by extending the surface distribution $f(s), 
s \in \Gamma$ to a function $f_{\Omega}(x)$ in $\Omega$ such that
\begin{equation}\label{eq:surf-delta}
f(s) = \int_{-L}^L f_{\Omega} (s+l \cdot \mathrm{n})\, \delta_\Gamma \; dl,
\end{equation}
where a surface delta function $\delta_{\Gamma}$ is introduced to restrict the volume concentration is essentially
concentrated in the neighborhood of width $2L$ symmetrically located on the surface $\Gamma$. There are many choices to use for the surface delta function.  
A catalog of possibilities is found in \cite{Lee2012}.  A good choice must maintain the property that $\int_\Omega \delta_\Gamma \, dx \propto$ area$(\Gamma)$ 
similar to the area constraint \eqref{eq:surf_constr}.  It is also convenient to choose a function without the use of $|\nabla \phi|$ for the 
sake of the simplicity of the variation computations with respect to $\phi$ (which is useful for computing the shape 
equation of the membrane \cite{Du2006}).  We choose the following function for numerical reasons,
\begin{equation}\label{eq:delta-new}
\delta_\Gamma = \left\{ \begin{split}
&\tanh(10(\phi+1)),   \quad &-1\leq\phi\leq0; \\
&-\tanh(10(\phi-1)),  \quad & 0\leq\phi\leq1. \\
\end{split} \right.
\end{equation}
Note that this function is continuous at $\phi = 0$; however, it does not have a continuous derivative.  But, the 
effect is negligible, since $\sech^2(D) \to -\sech^2(-D)$ as $D\to \infty$, and $D=10$ is large enough to avoid numerical problems in continuity.  


%
%
%
The Eulerian formulation of the membrane curvature energy involves the concentrations of lipids and proteins so it is mathematically 
possible to compute the variations of this energy with respect to these concentrations. This variation is defined everywhere in $\Omega$,
while it is the membrane curvature at the zero level set of the phase field function that is relevant to the molecular localization. 
Consequently, rather than computing a new variation directly from \eqref{eq:E-mem}, it is simpler to use the variations \eqref{eq:dELdp_lip}-
\eqref{eq:dELdp_pro} and replace the membrane mean curvature $H$ by a function of the phase field function. An $H$ expression  
consistent with the above definition of phase field function is given as \cite{Du2005} 
\begin{equation}\label{eq:H}
H = \frac{\sqrt{2}\epsilon}{2(\phi^2-1)}\left(\Delta\phi - \frac{1}{\epsilon^2} \phi(\phi^2-1) \right). 
\end{equation}
Again, this expression is only valid near the $\phi=0$ level set where the membrane surface is actually defined. Therefore, we restrict 
the mean curvature with a surface delta function so that it is only computed on (a narrow band around) the surface.  The 
computation of $H$ at other level sets of $\phi$ is extremely temperamental and introduces great numerical difficulties. Using this 
phase field approximation to the mean curvature in \eqref{eq:H} we will get an Eulerian formulation of the membrane curvature energy
with respect to the lipid concentrations: 
\begin{equation}\label{eq:mempart}
\frac{\delta E_{\rm mem}}{\delta \rho_l^{\rm lip}} \sim 2C_0^l (a_l^{\rm lip})^2 \left(C_0 - \frac{\varepsilon}{\sqrt{2}(\phi^2-1)} 
\left( \Delta \phi - \frac{1}{\varepsilon^2} \phi (\phi^2-1)  \right)\delta_\Gamma \right)  .
\end{equation}
Note that the diffusion potential for the lipids is defined over all of $\Omega$, but is nonzero only near the membrane $\Gamma$.  
The concentrations are initially distributed on the membrane only, and it is clear in \eqref{eq:flux-domain} below that the flux is restricted 
to the membrane only, hence these terms remain zero away from $\Gamma$.  This is also the case for the spontaneous curvature $C_0$, 
since it depends upon the concentrations. However, the mean curvature \eqref{eq:H} may be nonzero away from $\Gamma$, since this 
computation depends upon $\phi$, yet the expression is only relevant for the $\phi=0$ level set as we are using only the mean curvature
at $\phi=0$ to drive the transportation of lipids and proteins. Therefore, a surface delta function is applied to this term only to 
avoid irrelevant mean curvatures away from the $\phi=0$ level set.  

The full diffusion potential for the lipid species in the context of the phase field is now given by combining 
\eqref{eq:entropicpart} and \eqref{eq:mempart},
\begin{align}\label{eq:mu-lip}
\mu_l^{\rm lip} &= \frac{1}{\beta} \left[ \frac{-(a_l^{\rm lip})^2}{(a_0^{\rm lip})^2}  \ln\left(1-\rho^{\rm pro}(a^{\rm pro})^2-\sum_{j=1}^m \rho_j^{\rm lip} (a_j^{\rm lip})^2\right) + \ln(\rho_l^{\rm lip} (a_l^{\rm lip})^2) \right] \\
   & \qquad + 2C_0^l (a_l^{\rm lip})^2 \left(C_0 - \frac{\varepsilon}{\sqrt{2}(\phi^2-1)} \left( \Delta \phi - \frac{1}{\varepsilon^2} \phi (\phi^2-1)  \right)\delta_\Gamma \right) .\notag
\end{align}
The protein diffusion potential is computed similarly as
\begin{align}\label{eq:mu-pro}
\mu^{\rm pro} &=  \frac{1}{\beta} \Bigg[ \frac{-(a^{\rm pro})^2}{(a_0^{\rm lip})^2}  \ln\left(1-\rho^{\rm pro}(a^{\rm pro})^2-\sum_{j=1}^m \rho_j^{\rm lip} (a_j^{\rm lip})^2\right)  + \ln(\rho^{\rm pro} (a^{\rm pro})^2) \Bigg] \\
 & \quad + 2C_0^{\rm pro} (a^{\rm pro})^2 \left(C_0 - \frac{\varepsilon}{\sqrt{2}(\phi^2-1)} \left( \Delta \phi - \frac{1}{\varepsilon^2} \phi (\phi^2-1)  \right) \delta_\Gamma\right) \notag.
\end{align}


\subsection{Surface velocity}\label{sec:velcomp}
  
The fact that the surface deformation is driven by energy minimization implies that any component of the velocity that is 
tangential to the surface will cost extra energy for the deformation. In other words, a velocity field given by the
evolution of the phase field function and consistent with the minimization of the total energy must be normal to the 
zero level set of $\phi$.
With these justifications, we will derive a velocity field from the evolving phase field function using \cite{Osher2003}
\begin{equation}\label{eq:velocity-n}
\mathrm{v}_\mathrm{n} = \mathrm{n} \cdot \frac{dx}{dt} = \frac{\nabla \phi}{|\nabla \phi|} \frac{dx}{dt} =-\frac{\phi_t}{|\nabla \phi|}.
\end{equation}


\subsection{Curvature-driven transportation equation}
With the extension of the lipid and protein concentrations from the surface to domain, their transportation can not be described by the
surface equation \eqref{eq:mass-cons}. Rather, it will be governed by following general transportation equation defined in the entire $\Omega$:
\begin{equation}\label{eq:mass-cons-domain}
\frac{\partial \rho}{\partial t} + \mathrm{v} \cdot \nabla \rho = - \nabla \cdot \mathrm{J},
\end{equation}
with the flux vector being defined now by
\begin{equation}\label{eq:flux-domain}
\mathrm{J} = - D_{\Omega} \delta_{\Gamma} \beta \rho \nabla_s \mu,
\end{equation}
where $D_{\Omega}$ is the volume diffusion coefficient, which can be inversely determined by using the measured lateral diffusion coefficients 
of lipids or proteins and a relation modeling \eqref{eq:surf-delta}: $D_{\Gamma} = D_{\Omega} \int_{-L}^L \delta_\Gamma \; dl$.

We are now in a position to rearrange the diffusion equation \eqref{eq:mass-cons-domain} with individual terms computed above. 
We define
\begin{align}
&L^{\rm lip}(\rho_l^{\rm lip}) = \ln(\rho_l^{\rm lip} (a_l^{\rm lip})^2), \label{eq:lead-lip} \\
&R^{\rm lip}(\rho_l^{\rm lip}, \rho^{\rm pro}) = \frac{-(a_l^{\rm lip})^2}{(a_0^{\rm lip})^2}  \ln\left(1-\rho^{\rm pro}(a^{\rm pro})^2-\sum_{j=1}^m \rho_j^{\rm lip} (a_j^{\rm lip})^2\right) , \label{eq:remainder-lip} \displaybreak[0]\\
&L^{\rm pro}(\rho^{\rm pro}) = \ln(\rho^{\rm pro} (a^{\rm pro})^2), \label{eq:lead-pro} \\
&R^{\rm pro}(\rho_l^{\rm lip}, \rho^{\rm pro}) = \frac{-(a^{\rm pro})^2}{(a_0^{\rm lip})^2}  \ln\left(1-\rho^{\rm pro}(a^{\rm pro})^2-\sum_{j=1}^m \rho_j^{\rm lip} (a_j^{\rm lip})^2\right) , \label{eq:log-term-pro} \displaybreak[0]\\
&P(\phi,\rho_l^{\rm lip}, \rho^{\rm pro}) = \left(C_0(\rho_l^{\rm lip}, \rho^{\rm pro}) - \frac{\varepsilon}{\sqrt{2}(\phi^2-1)} \left( \Delta \phi - \frac{1}{\varepsilon^2} \phi (\phi^2-1) \right) \delta_\Gamma\right) . \label{eq:remainder-phi}
\end{align}
The notation is indicative of leading order terms for the lipids and proteins ($L$), remaining terms corresponding to the size restrictions ($R$), 
and a term corresponding to the curvature determined by the phase field function ($P$).  Using \eqref{eq:lead-lip}-\eqref{eq:remainder-phi} and 
suppressing the notation describing each function's independent variables, the diffusion potentials \eqref{eq:mu-lip} and \eqref{eq:mu-pro} become
\begin{align}
\mu_l^{\rm lip} &= \frac{1}{\beta} (L^{\rm lip} + R^{\rm lip})  + 2 C_0^l (a_l^{\rm lip})^2 P , \label{eq:mu-lip-LP} \\
\mu^{\rm pro} &= \frac{1}{\beta}  (L^{\rm pro} + R^{\rm pro}) + 2 C_0^{\rm pro} (a^{\rm pro})^2 P . \label{eq:mu-pro-LP}
\end{align}
Then, to compute the flux \eqref{eq:flux-domain} we have 
\begin{align}
\nabla \mu_l^{\rm lip} &=  \frac{1}{\beta} (\nabla L^{\rm lip} + \nabla R^{\rm lip}) +  2 C_0^l (a_l^{\rm lip})^2 \nabla P,  \label{eq:d-mu-lip} \\
\nabla \mu^{\rm pro} &=  \frac{1}{\beta} ( \nabla L^{\rm pro} +\nabla R^{\rm pro} )  +  2 C_0^{\rm pro} (a^{\rm pro})^2 \nabla P, \label{eq:d-mu-pro}
\end{align}
where we note the derivatives of the leading terms are simply
\begin{align}
&\nabla L^{\rm lip} = \frac{\nabla \rho_l^{\rm lip}}{\rho_l^{\rm lip} }, \label{eq:d-lead-lip}\\
&\nabla L^{\rm pro} = \frac{\nabla \rho^{\rm pro}}{\rho^{\rm pro} }. \label{eq:d-lead-pro} 
\end{align}
Also define
\begin{align}
&M_l^{\rm lip} = 2 \beta C_0^l (a_l^{\rm lip})^2,  \label{eq:Mlip} \\
&M^{\rm pro} = 2 \beta C_0^{\rm pro} (a^{\rm pro})^2.  \label{eq:Mpro}
\end{align}
%
Using these variables, the diffusion equation \eqref{eq:mass-cons-domain} for each species becomes
\begin{align}
\frac{\partial  \rho^{\rm lip}_l}{\partial t} + \mathrm{v} \cdot \nabla \rho_l^{\rm lip} &= D_l^{\rm lip} \nabla \cdot \Bigg\{ \delta_\Gamma \nabla \rho_l^{\rm lip}  + \delta_\Gamma \rho_l^{\rm lip} \nabla R^{\rm lip} + M_l^{\rm lip} \delta_\Gamma \rho_l^{\rm lip} \nabla P \Bigg\},  \label{eq:mass-cons-lip}\\
\frac{\partial  \rho^{\rm pro}}{\partial t} + \mathrm{v} \cdot \nabla \rho^{\rm pro} &= D^{\rm pro} \nabla \cdot \Bigg\{ \delta_\Gamma \nabla \rho^{\rm pro}  + \delta_\Gamma \rho^{\rm pro} \nabla R^{\rm pro} + M^{\rm pro} \delta_\Gamma \rho^{\rm pro} \nabla P \Bigg\}. \label{eq:mass-cons-pro}
\end{align}


%
%
%
%
The equations above take the form of the drift-diffusion equation.  
If we neglect the size effect terms involving $R$, and consider the equation on a stationary membrane ($\mathrm{v}=0$), the equations take the form
\begin{equation}\label{eq:drift-diffusion0}
\frac{\partial \rho}{\partial t} = D \nabla \cdot (\delta_\Gamma \nabla \rho + M \delta_\Gamma \rho \nabla(C_0-H\delta_\Gamma)).
\end{equation}
If we use the surface operator, the use of delta functions can be absorbed, giving rise to
\begin{equation}\label{eq:drift-diffusion}
\frac{\partial \rho}{\partial t} = D \nabla_s \cdot (\nabla_s \rho + M \rho \nabla_s (C_0-H)).
\end{equation}
The second term in the equation shows that the drift of the diffusive species is due to the difference in the actual 
membrane curvature and the spontaneous curvature of the membrane $C_0-H$. It is this term that drives the localization
of lipids and proteins to the position on membrane surface where the preferred mean curvature is observed.

\subsection{Analytical solution to steady state surface diffusion}\label{sec:drift-diffusion}

The steady state drift-diffusion equation has an analytical solution.  
Consider the change of variables $ u = \rho e^{M(C_0-H)}$ \cite{Slotboom1973}.  The steady state form of \eqref{eq:drift-diffusion} takes an 
equivalent, symmetric form,
\begin{equation}\label{eq:drift-diffusion_slotboom}
0 = \nabla_s \cdot (e^{-M(C_0-H)} \nabla_s u ).
\end{equation}
It is straightforward to show that a solution to \eqref{eq:drift-diffusion_slotboom} is given by $u=c$ for some constant $c$.  From this, we have
\begin{equation}\label{eq:drift-diffusion_solution}
c = \rho e^{M(C_0-H)}.
\end{equation}
Solving \eqref{eq:drift-diffusion_solution} for $\rho$ gives
\begin{equation}\label{eq:dd-sol}
\rho = c e^{-M(C_0-H)}. 
\end{equation}
The integration of the lipid concentration over the membrane $\Gamma$ gives the total number of lipids on a leaflet, a conserved quantity,  
\begin{equation}\label{eq:T}
T = \DD\int_\Gamma \rho \; dS = \int_\Gamma c e^{-M(C_0-H)} \; dS.
\end{equation}
Factor the constant $c$ out of the integral and solve to get
\begin{equation}\label{eq:constant}
c = \frac{T}{\DD\int_\Gamma e^{-M(C_0-H)} \; dS}.
\end{equation}
Plugging \eqref{eq:constant} back into \eqref{eq:dd-sol}, we have the analytical solution to the steady state of \eqref{eq:drift-diffusion},
\begin{equation}\label{eq:rho}
\rho[\Gamma] = \frac{T e^{-M(C_0-H)}}{\DD\int_\Gamma e^{-M(C_0-H)} \; dS}.
\end{equation}

\section{Computational Simulations}\label{sec:results}


In the computational simulations, we solve the diffusion equations \eqref{eq:drift-diffusion} numerically for two competing species, one diffusive proteins, 
the other background lipids. By using a two-species model, we only need to solve one governing equation, since the concentration of the 
second (background) species can be computed directly from the concentration of the diffusive species according to 
saturation condition \eqref{eq:lipid-cons}. To further focus on the curvature induced molecular localization, we neglect the correction terms 
in the equation accounting for the size effects of molecules ($R$ in \eqref{eq:remainder-lip}).  
We use the Fourier spectral method to solve the equation in a similar fashion to the shape equation solution outlined in \cite{Du2006}.  
The detailed numerical procedure will be described in a future article. Throughout the numerical results, we consistently 
choose $\epsilon = 0.1$. 
We compare the results of pure diffusion without any curvature effects (by neglecting the $M\rho\nabla_s(C_0-H)$ term) to 
the diffusion with the curvature effects.  


We choose a torus as a test surface because it has regions of positive curvature and regions of negative curvature which 
we can analytically compute to test the curvature effects.  The surface is defined by 
\begin{equation}\label{eq:torus}
\left( R- \sqrt{x^2+y^2} \right)^2 + z^2 = r^2,
\end{equation}
where $R$ and $r$ are the major and minor radii, respectively.  The major radius $R$ is the distance from the center of the tube 
to the center of the torus, and the minor radius $r$ is the radius of the tube.  In this definition, the torus' hole 
is located along the $z$ axis.  The torus may be parameterized by $\theta$ and $\phi$, 
\begin{equation}\label{eq:torus-param}
\vecx = \begin{pmatrix}
(R+r\cos\theta)\cos\phi \\
(R+r\cos\theta)\sin\phi \\
r\sin\phi
\end{pmatrix},
\end{equation}
where the parameters $0 \leq \theta, \phi \leq 2\pi$.  The angle $\phi$ is the angle made from the 
surface to the positive $x$-axis (projected on the $xy$-plane), known as the toroidal angle, and the angle $\theta$ is the 
angle made from the surface around the center of the tube, known as the poloidal angle.  We consider a ring torus, where $R > r$.  
%
We choose $R = 2.0$ and $r = 1.1$, and solve the diffusion equation over a grid of $[-4,4]^3$, with a $128\times128\times128$ mesh.  

The initial concentration of diffusive proteins is chosen to be localized along the ring around the positive $x$-axis, smoothly 
distributed along the surface and smoothly distributed from the surface to the bulk,
\begin{align}
\rho(x,y,z,0) &= S \cdot \rm{exp}\left( -\sqrt{\left(x-R\right)^2+y^2+z^2} \right) \notag \\
              & \quad \cdot \rm{exp}\left( -2\left(r-\sqrt{(x-c_x)^2+(y-c_y)^2 + z^2}\right) \right), \label{eq:conc_init-torus}
\end{align}
%
where the point $(c_x, c_y, 0)$ is the center of the torus tube at a given angle $\phi$, and $S$ is chosen so 
that the maximum of the concentration is 1 on the torus surface.  

\begin{figure}[!ht]
   \centering
   \includegraphics[width=6cm]{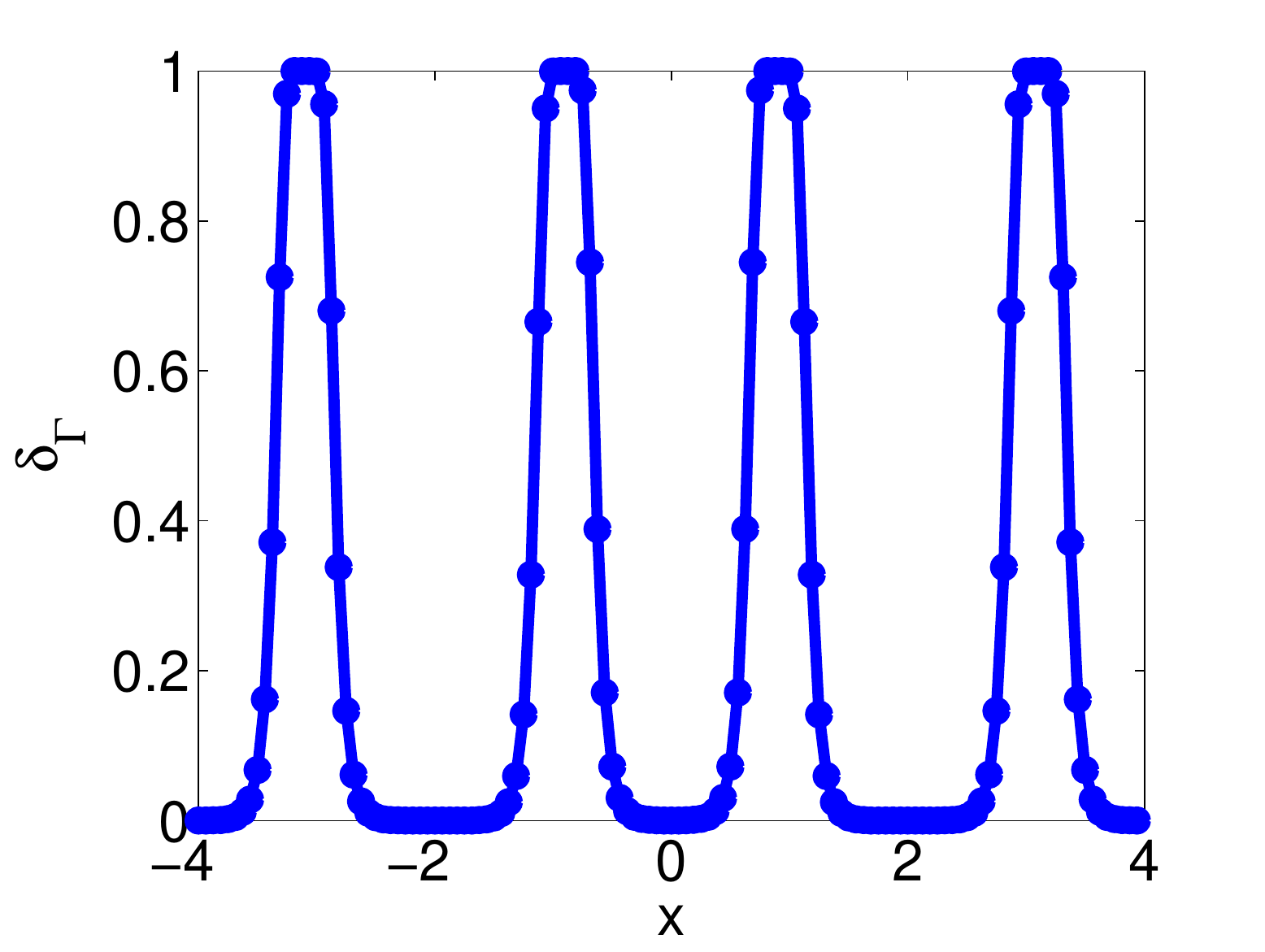} 
   \includegraphics[width=6.5cm]{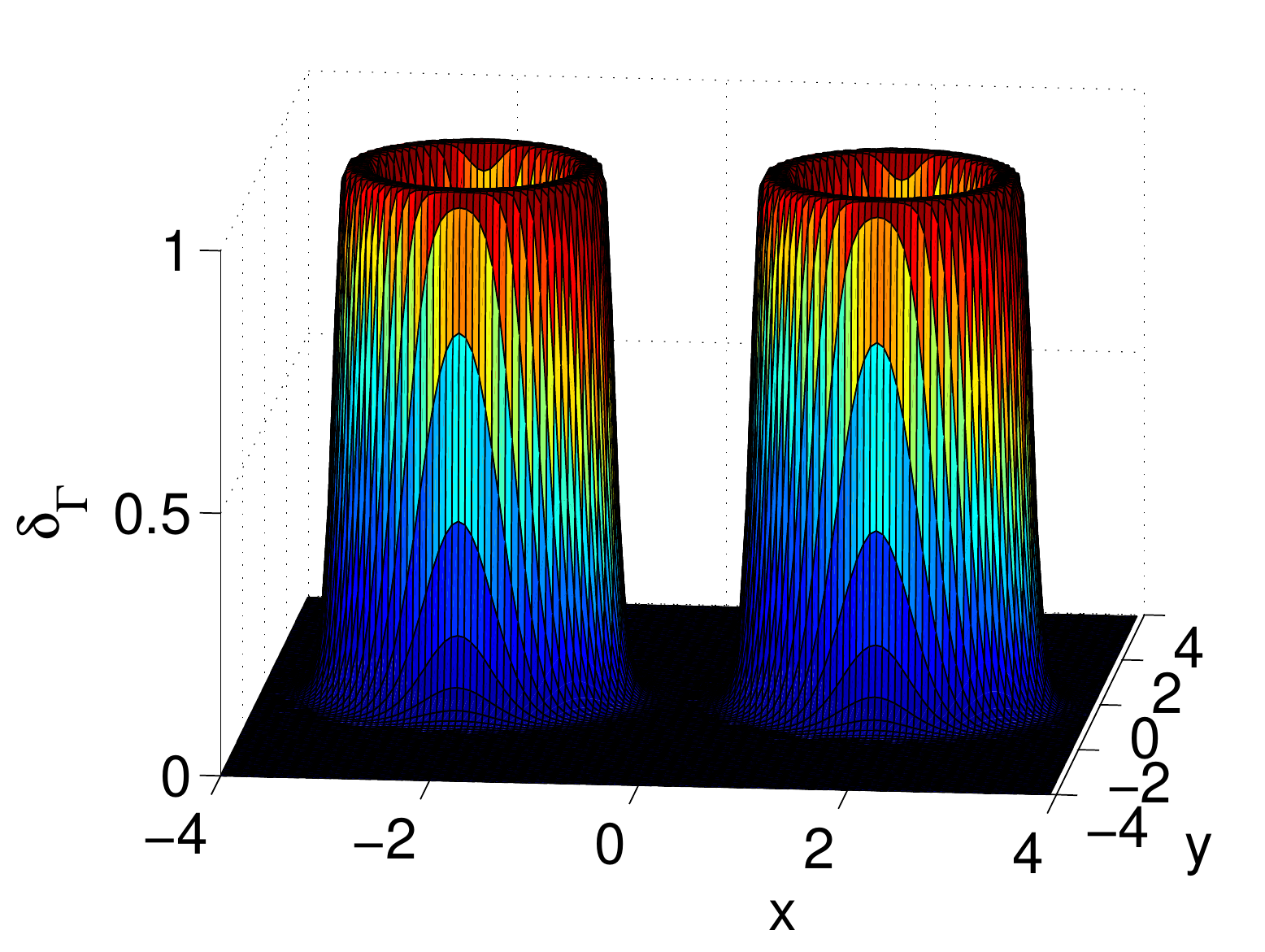} 
   \caption{Profile of delta function for the torus with radii $R=2.0$ and $r=1.1$ under a $128\times128\times128$ mesh.  
Left: 1D profile ($y=z=0$) of delta function across the $x$ grid points.  Right: 2D profile ($y=0$) across $x$ and $z$ grid points.} 
   \label{fig:delta-prof-torus2_128} 
\end{figure}

The profile of the delta function \eqref{eq:delta-new} for the torus is shown in Figure \ref{fig:delta-prof-torus2_128}. 
The 1D cross section of the delta function as seen in Figure \ref{fig:delta-prof-torus2_128} shows that the surface is well resolved over 
the $y=z=0$ cross section.  That is, there is a clear transition from the outside to the inside of each ring, and the delta function 
settles to zero in the empty space.  

\begin{figure}[!ht]
   \centering
 \begin{tabular}{lll}
 \includegraphics[width=3.125cm]{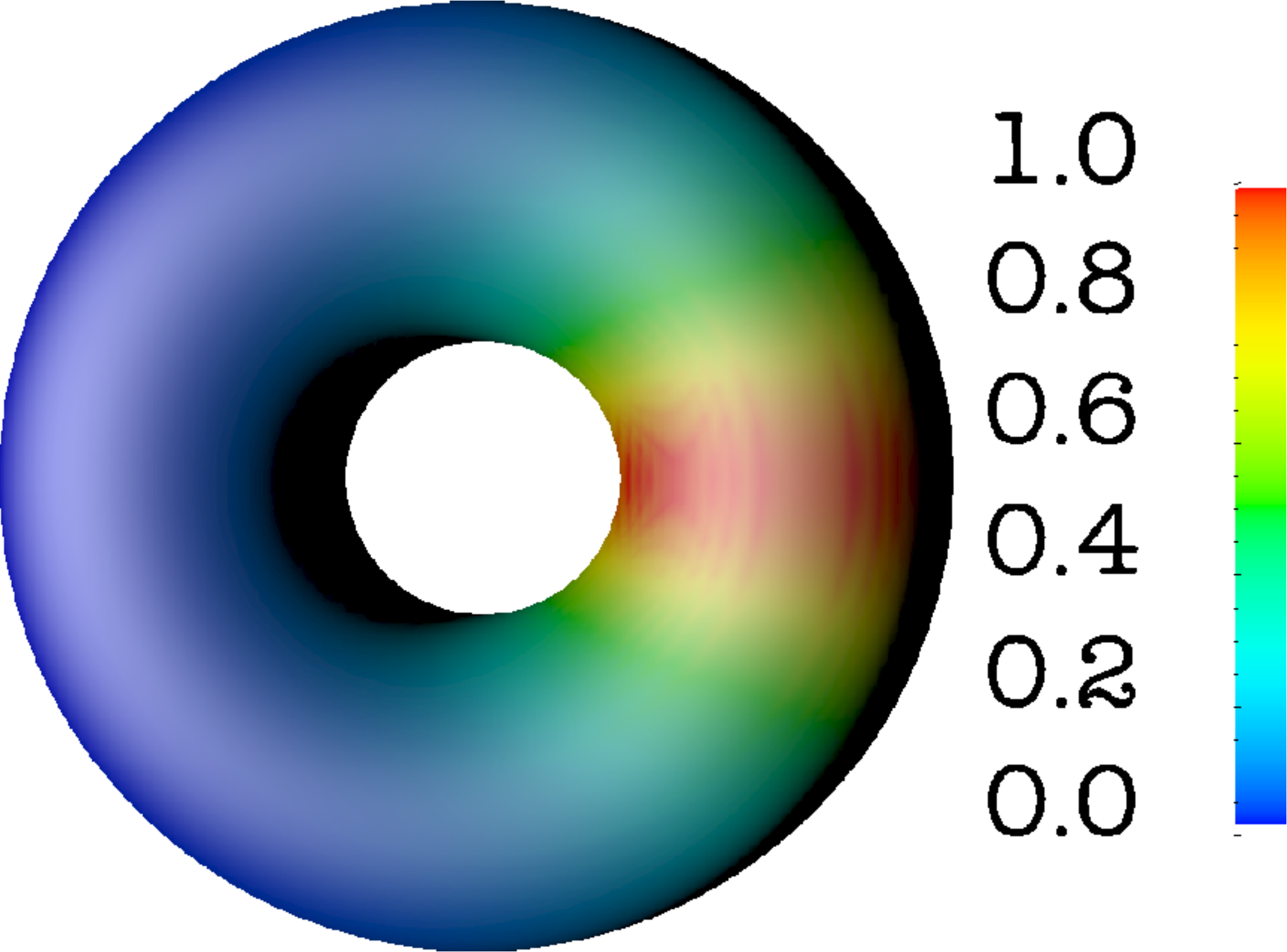}   &  
 \includegraphics[width=3.125cm]{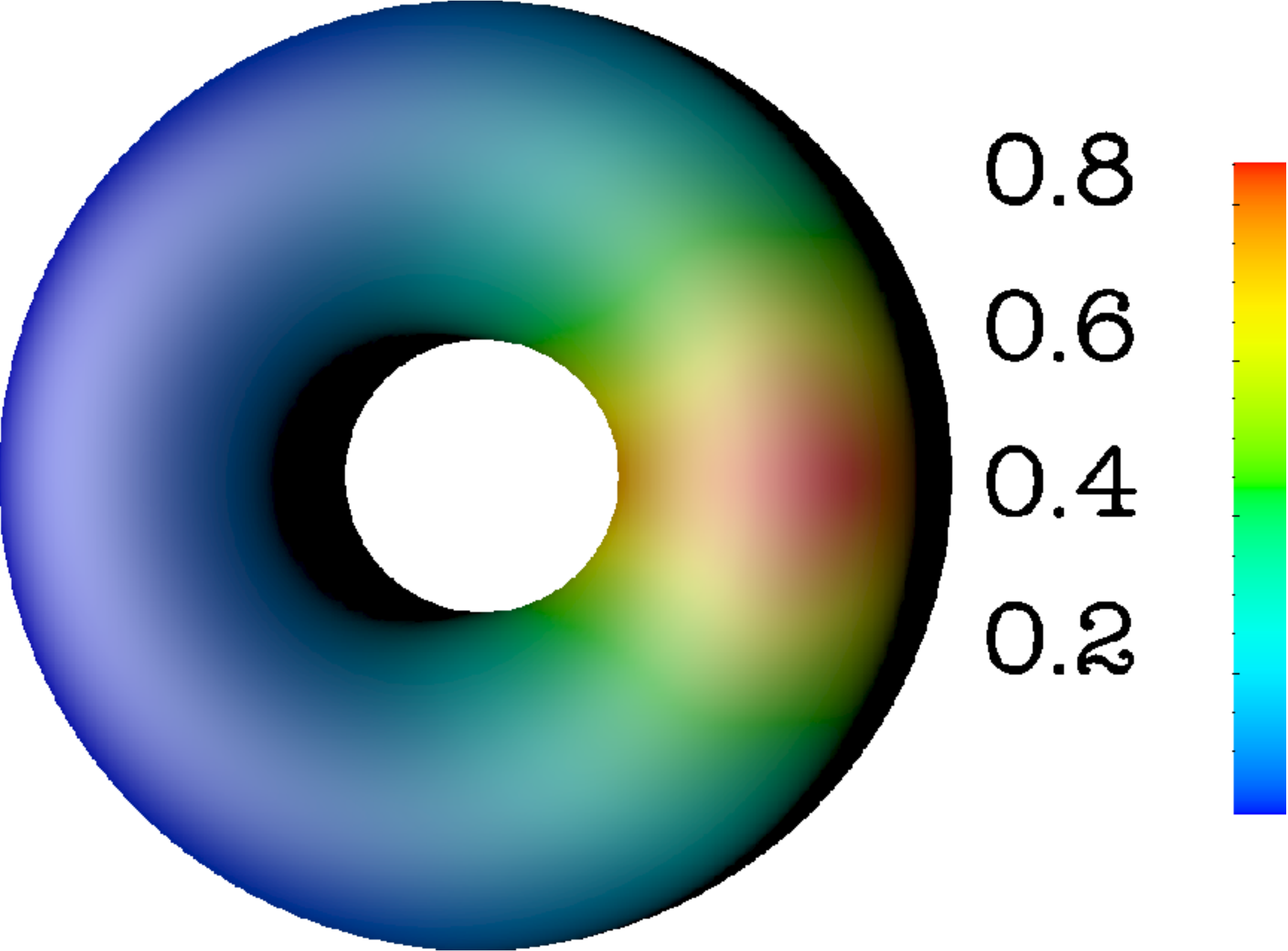} &
 \includegraphics[width=3.125cm]{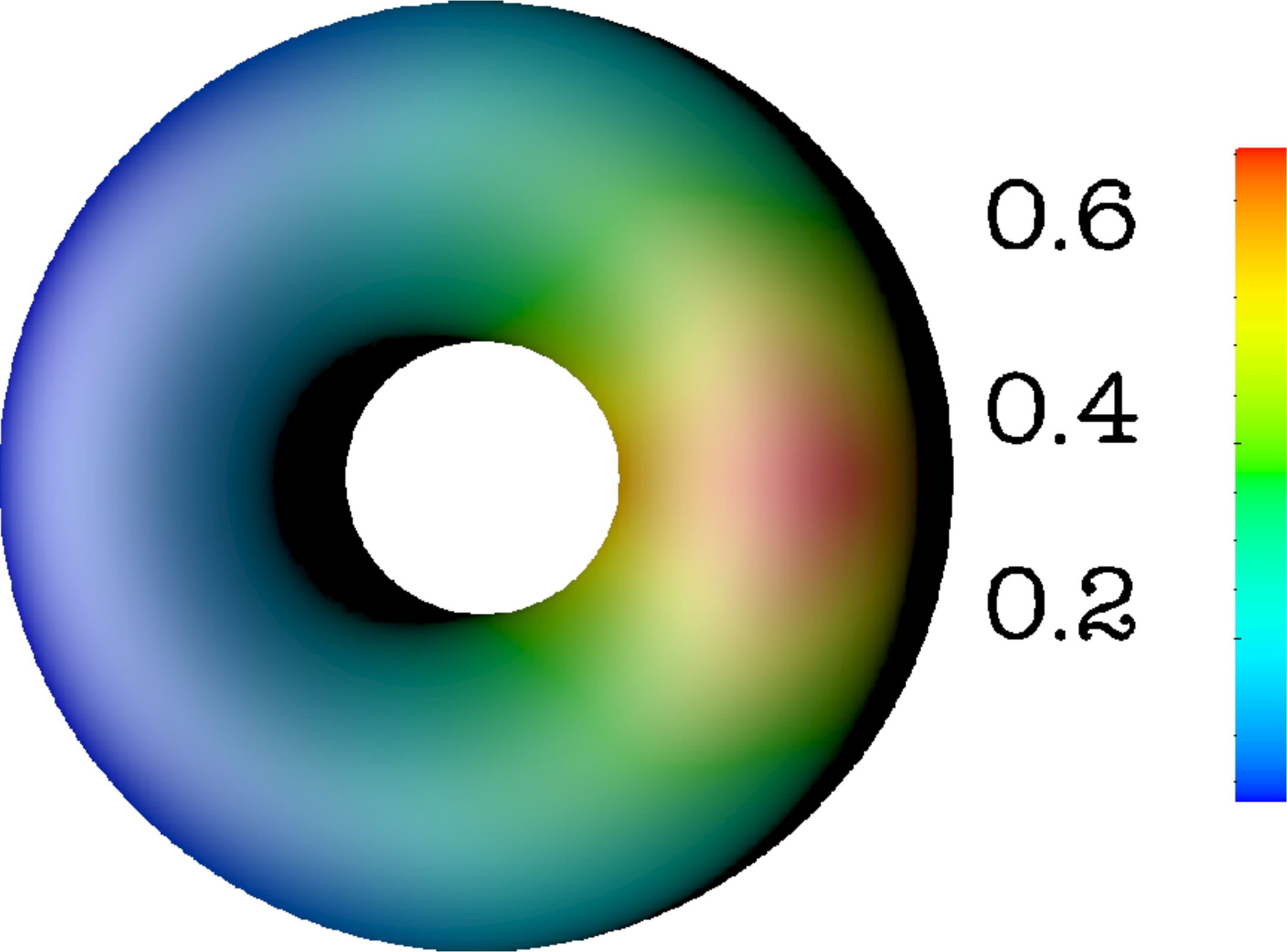} \\
 \includegraphics[width=3.125cm]{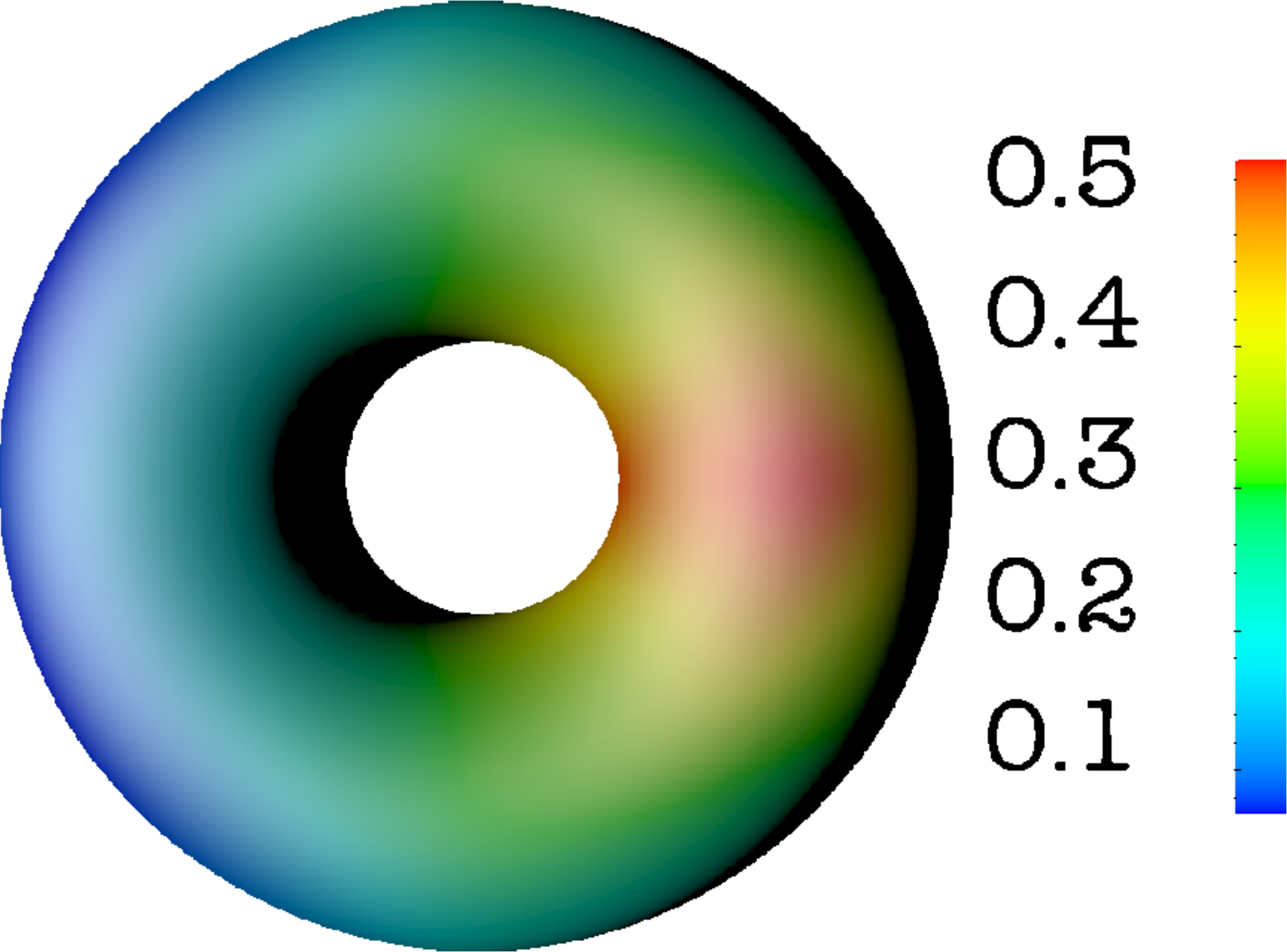} &
 \includegraphics[width=3.125cm]{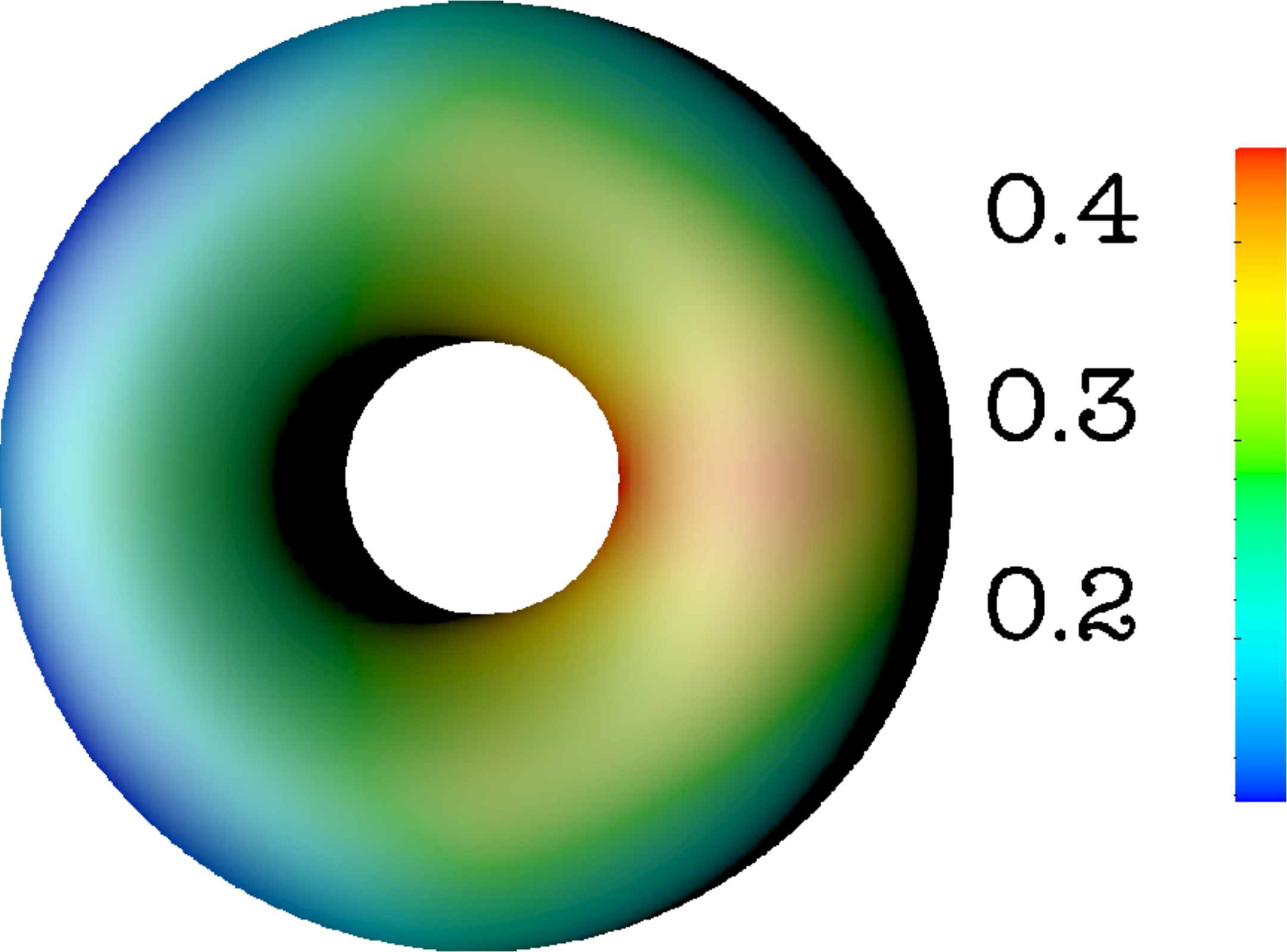} &
 \includegraphics[width=3.125cm]{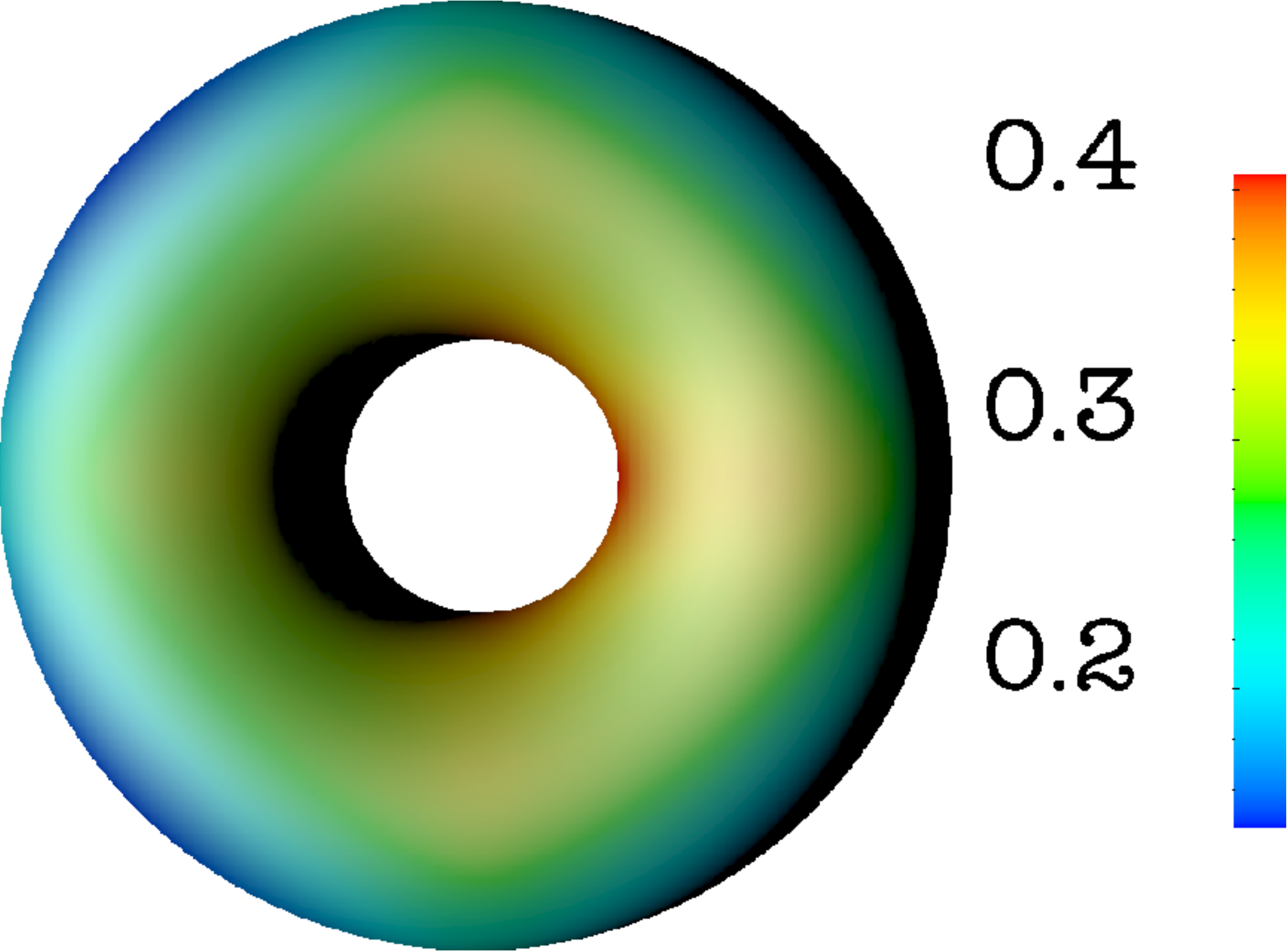} \\
 & & \\
 \includegraphics[width=2.75cm]{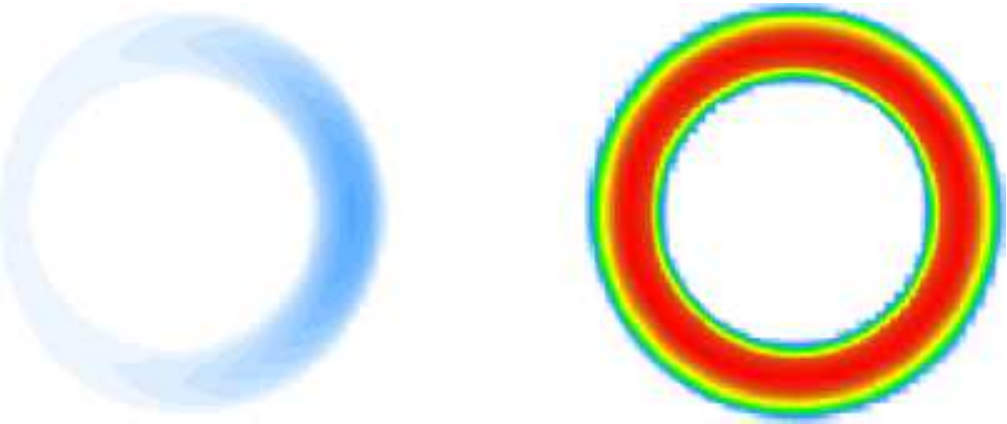}   &
 \includegraphics[width=2.75cm]{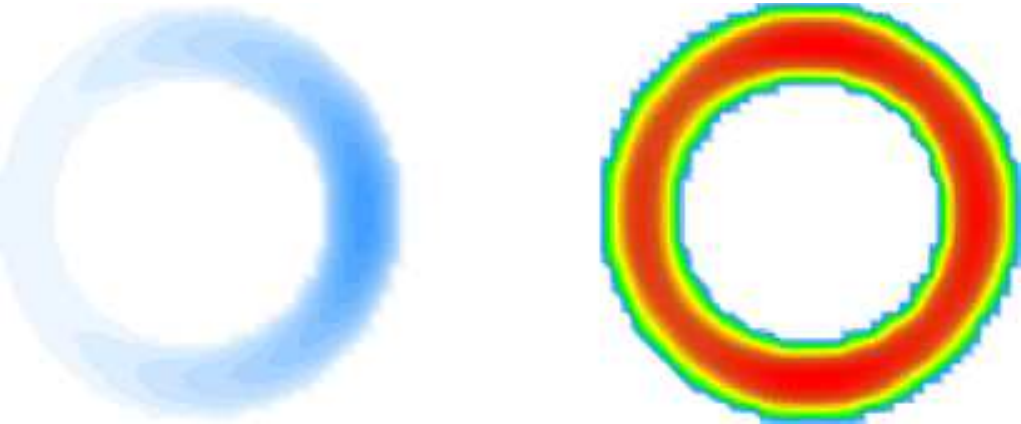} &
 \includegraphics[width=2.75cm]{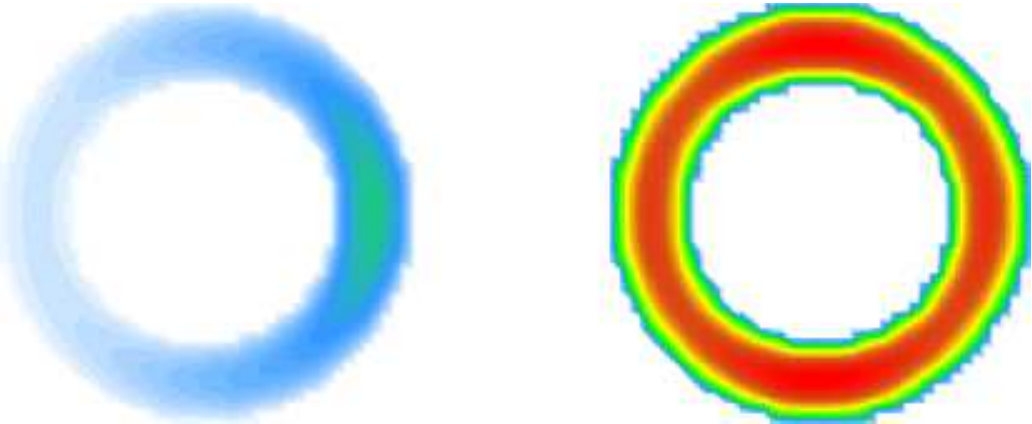} \\
 & & \\
 \includegraphics[width=2.75cm]{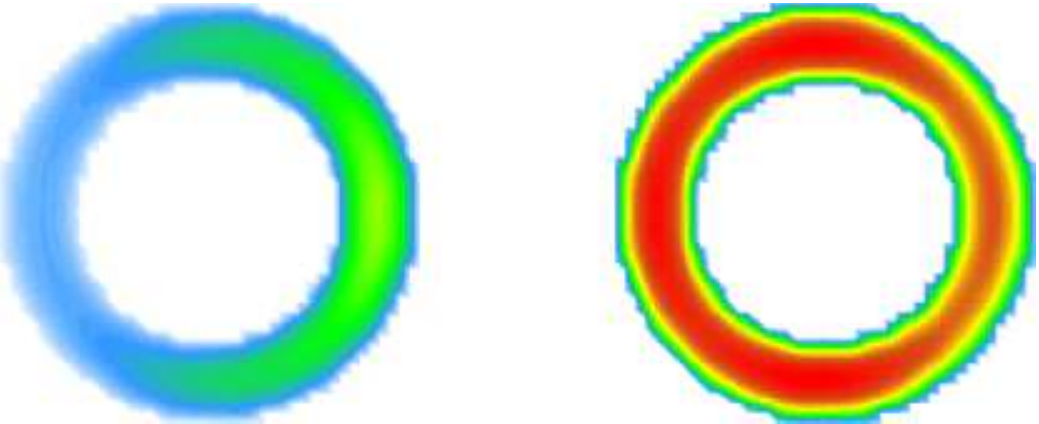} &
 \includegraphics[width=2.75cm]{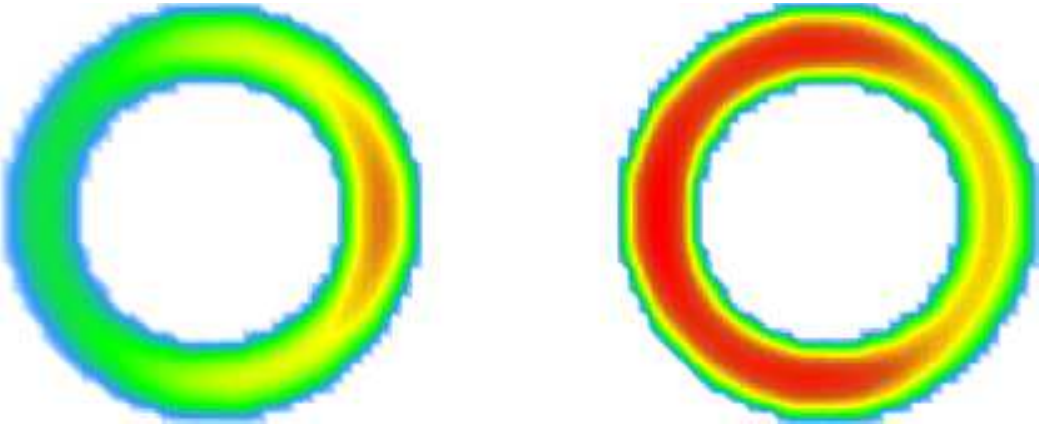} &
 \includegraphics[width=2.75cm]{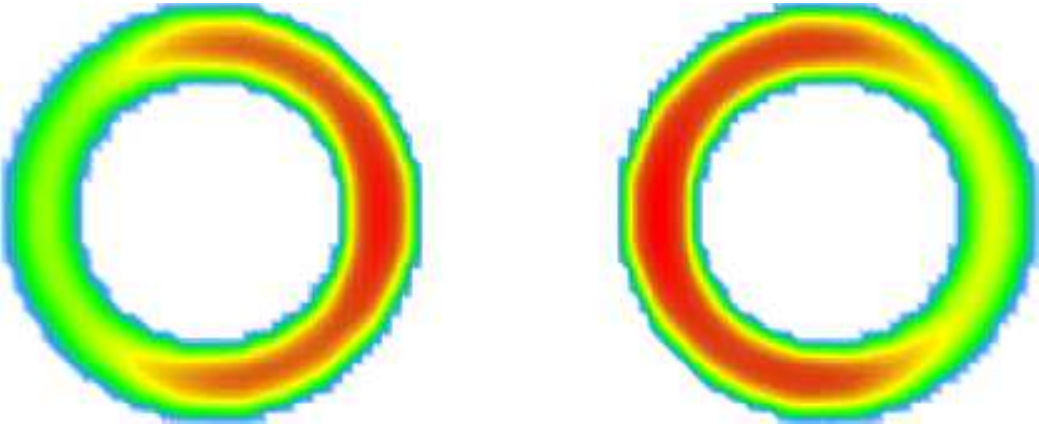}  
 \end{tabular} 
   \caption{Concentration of diffusive proteins on surface and on a 2D cross section of the torus, governed by pure 
diffusion with $128\times128\times128$ mesh.  Times shown are $t = 0, 0.25, 1.0, 2.5, 5.0$, and $10.0$. The color is scaled by the maximum 
concentration in each plot.}
   \label{fig:torus2_N128} 
\end{figure}

The results of pure diffusion neglecting curvature effects with the new mesh are presented in Figure \ref{fig:torus2_N128}.
Notice that the numerical simulation did not resolve the diffusion equation well, since the grid spacing is $\Delta x = 8/128 = 0.0625$.  
This is poor resolution, yet we do not increase the number of grid points since the computation of the 3D problem 
in the phase field context is expensive.  In spite of the poor resolution, curvature effects are still clearly demonstrable and are shown next.

\begin{figure}[!ht]
   \centering
   \includegraphics[width=7cm]{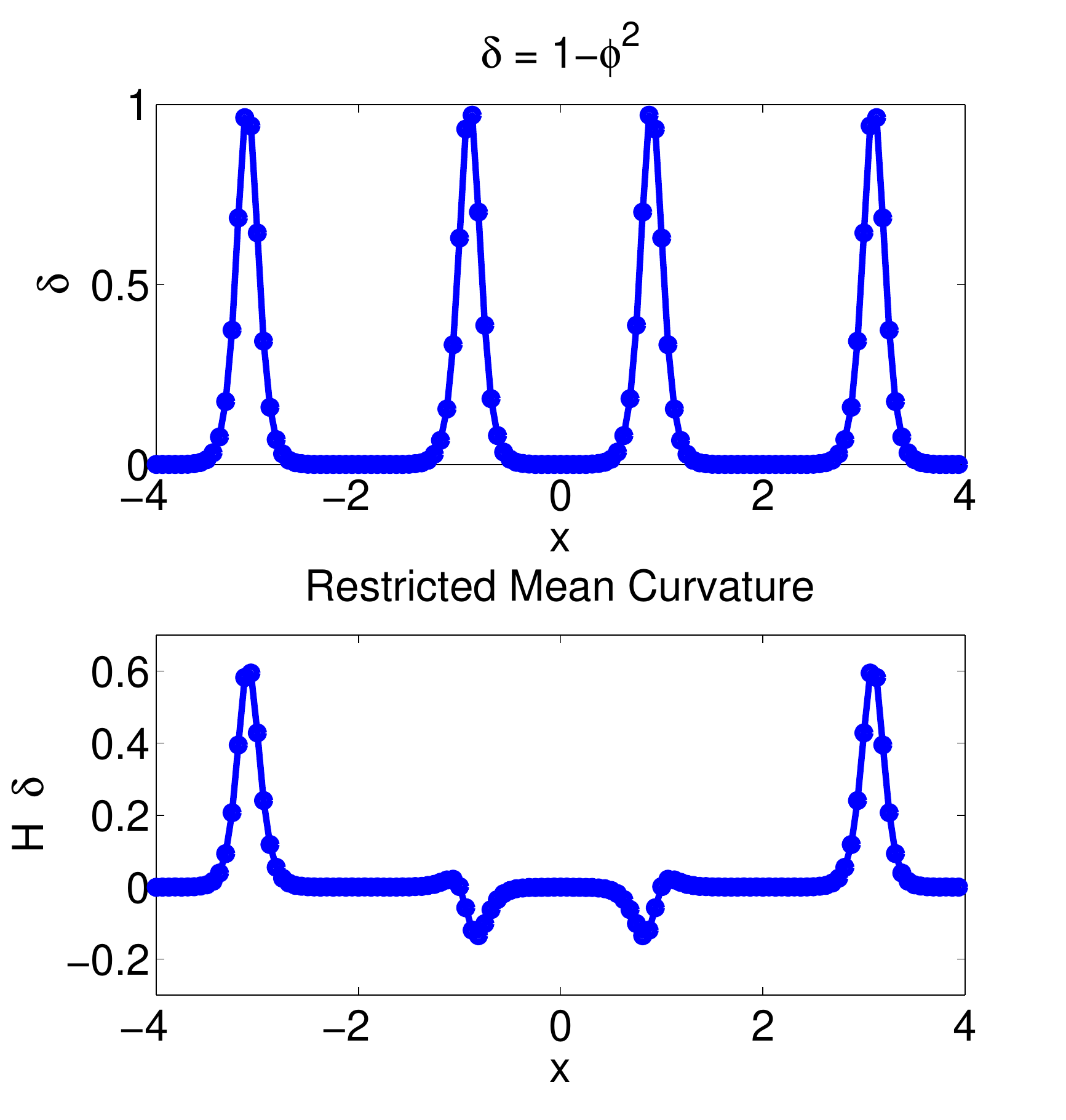} 
   \caption{Mean curvature of a torus with $128\times128\times128$ mesh.  Top: 1D profile ($y=z=0$) of a (simpler) delta function used to restrict 
mean curvature \eqref{eq:H} to the surface.  Bottom: 1D profile of numerical mean curvature of torus.  Note the 
match to the analytical values of 0.6 and -0.1.} 
   \label{fig:H_torus_128} 
\end{figure}

Analytically, the mean curvature of a torus is given by
\begin{equation}\label{H-torus}
H_{torus} = \frac{R+2r\cos\theta}{2r(R+r\cos\theta)} .
\end{equation}
At the outer ring of the torus, $\theta = 0$.  Therefore the mean curvature is $H_{torus} \approx 0.6158$.   
At the inner ring of the torus, $\theta = \pi$, and $H_{torus} \approx -0.1010$.  
A 1D cross section of the numerical mean curvature is shown in Figure \ref{fig:H_torus_128}.  
The plot matches the analytical values at the positions of the inner and outer rings well. 

\begin{figure}[!ht]
   \centering
\begin{tabular}{lll}
 \includegraphics[width=3.125cm]{Torus_N128_t0B}   &  
 \includegraphics[width=3.125cm]{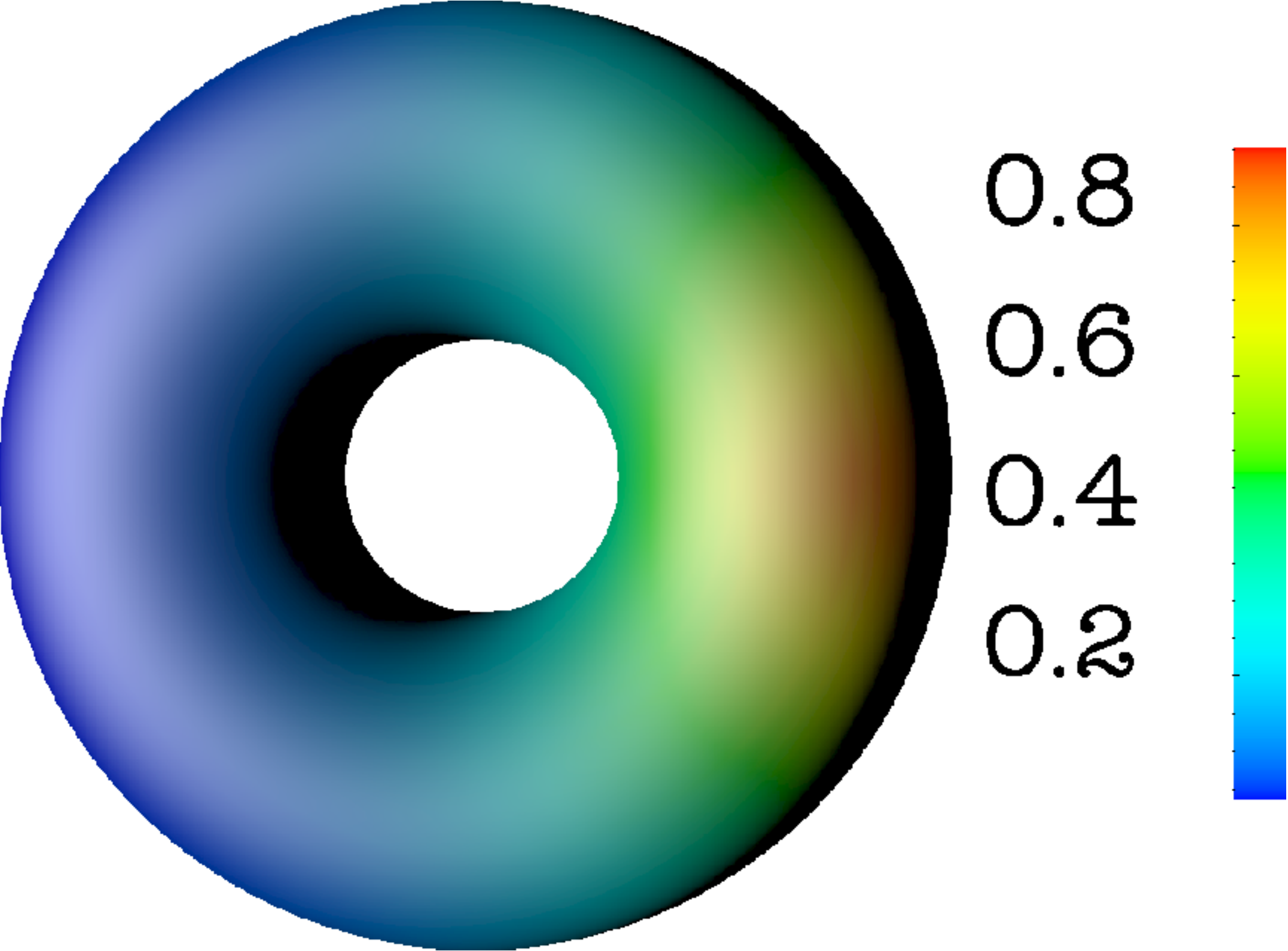} &
 \includegraphics[width=3.125cm]{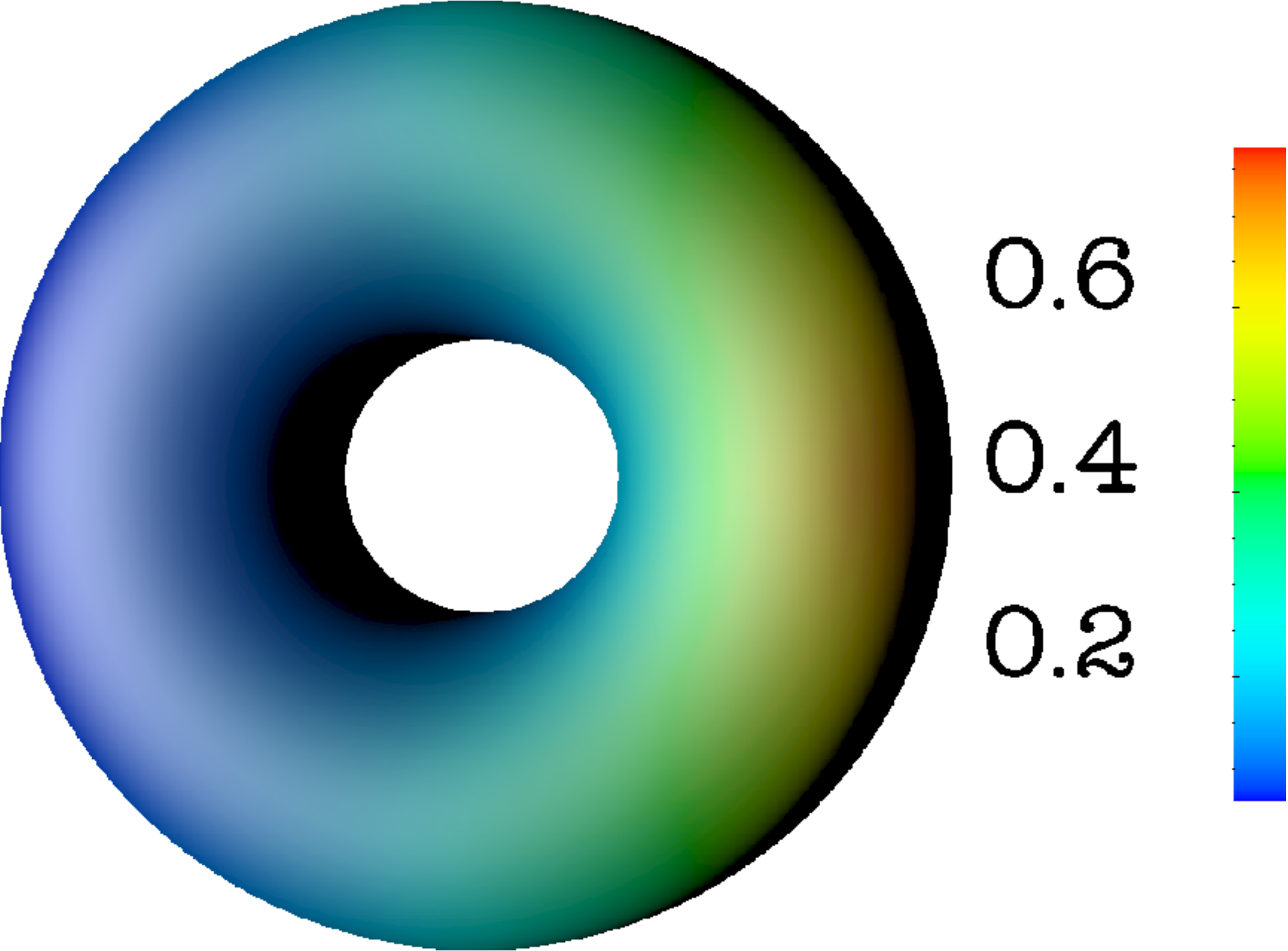} \\ 
 \includegraphics[width=3.125cm]{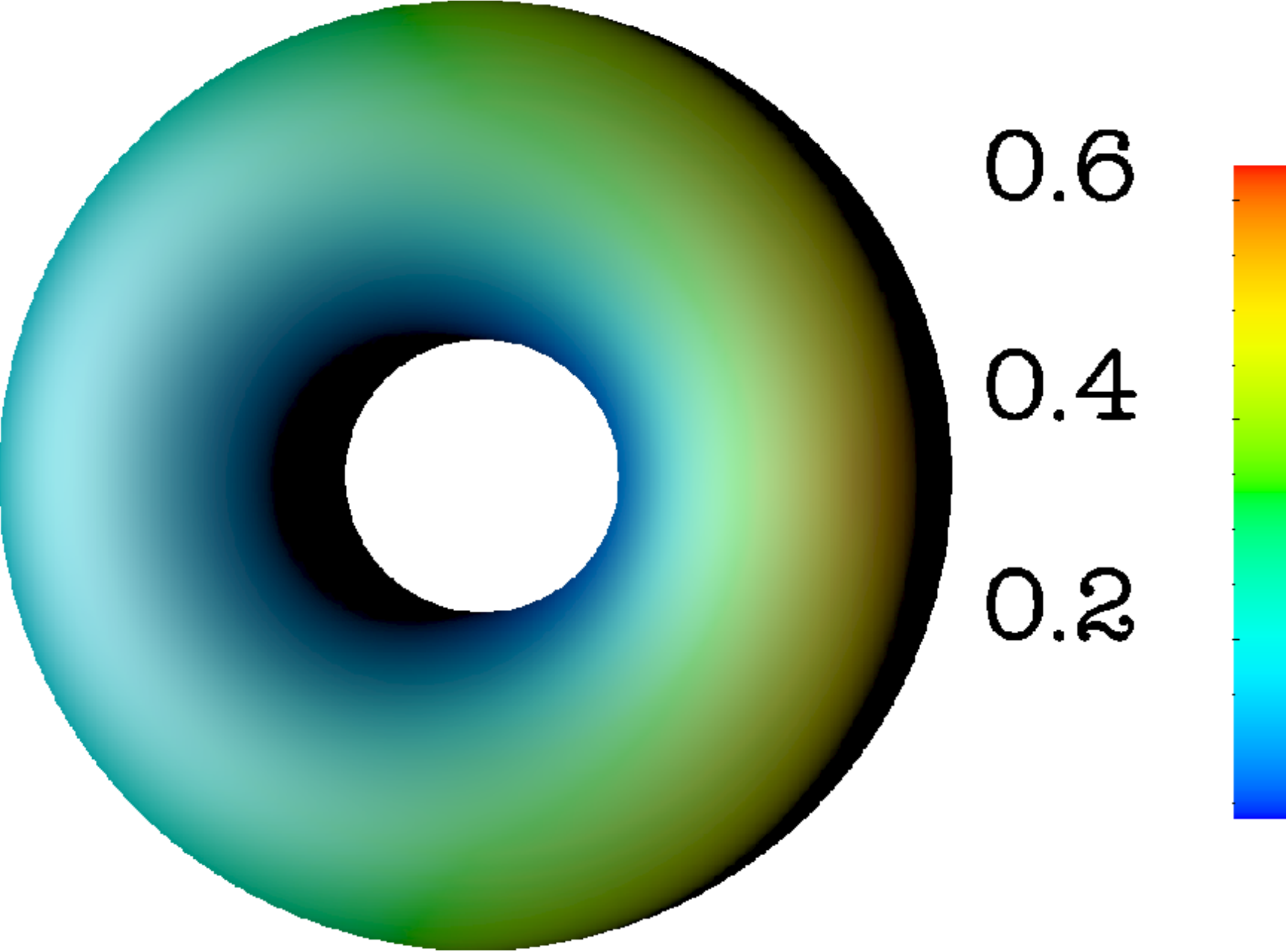} &
 \includegraphics[width=3.125cm]{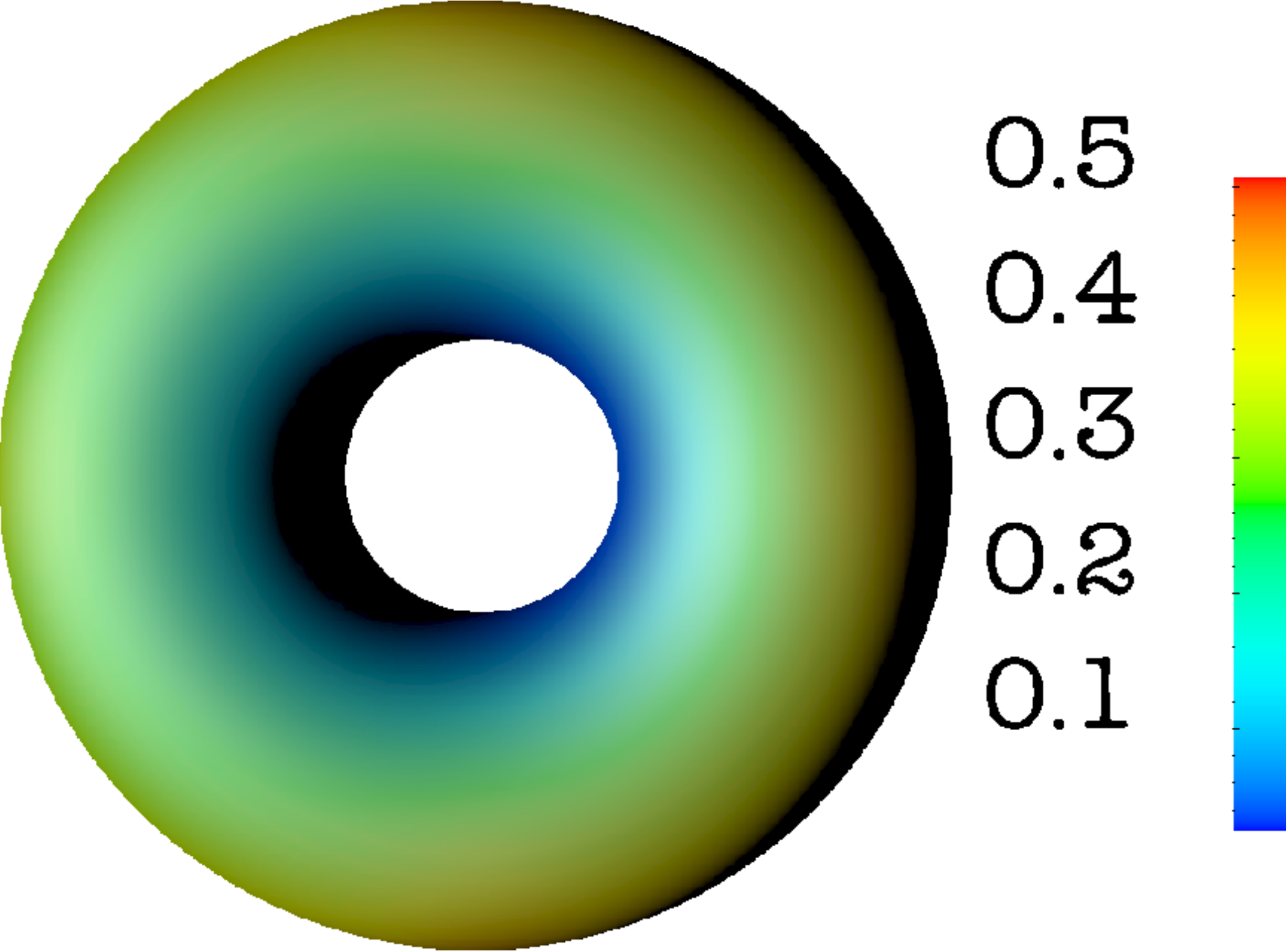} &
 \includegraphics[width=3.125cm]{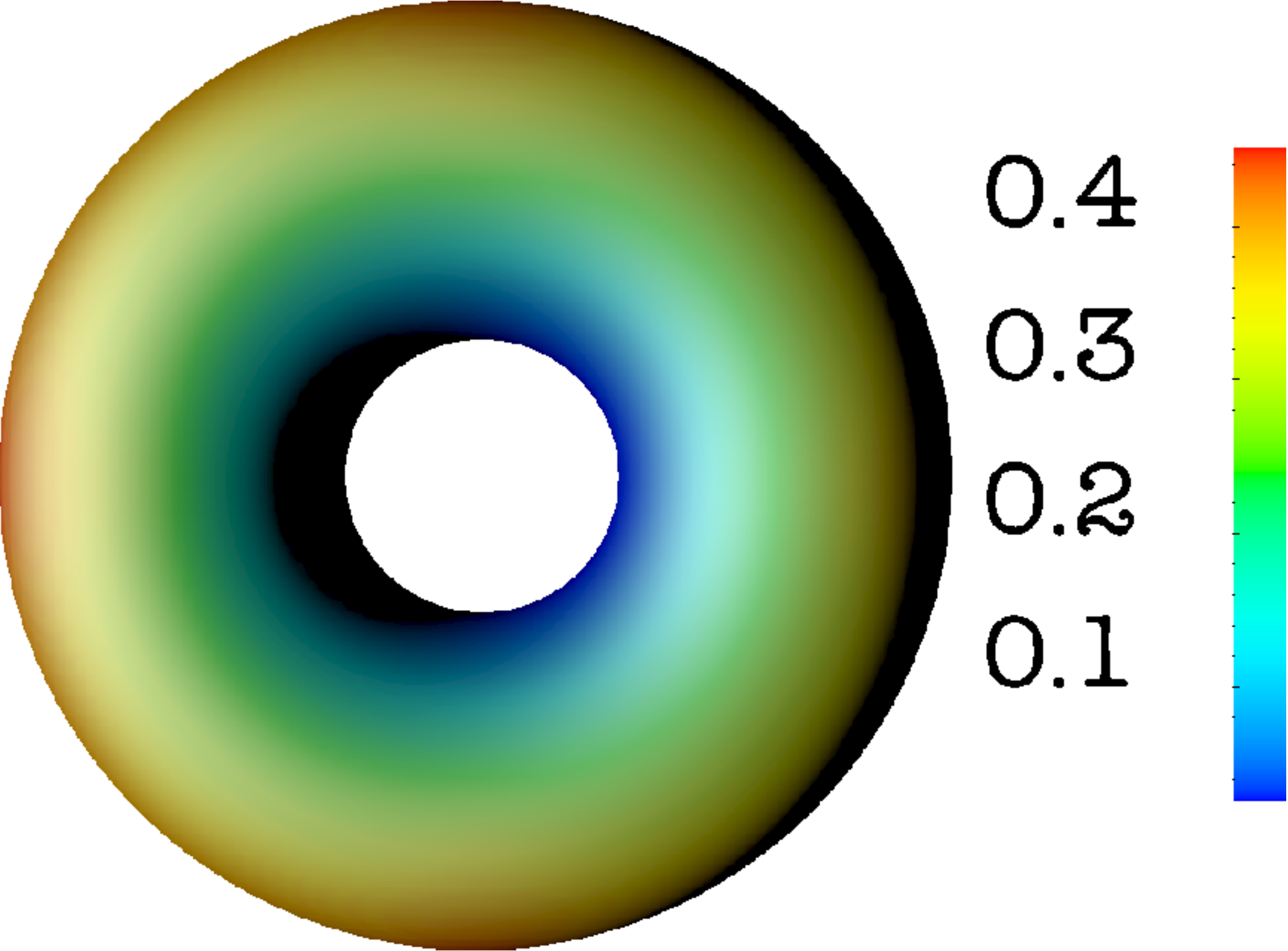} \\ 
  & & \\
 \includegraphics[width=2.75cm]{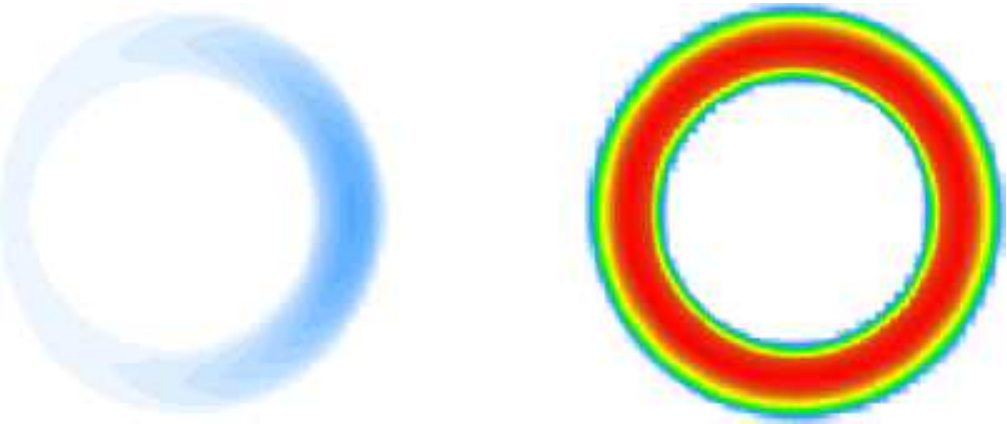}   &  
 \includegraphics[width=2.75cm]{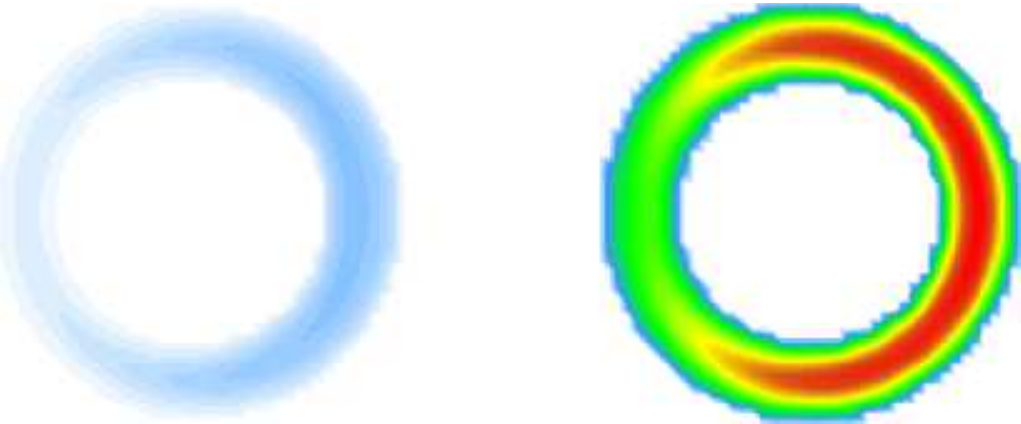} &
 \includegraphics[width=2.75cm]{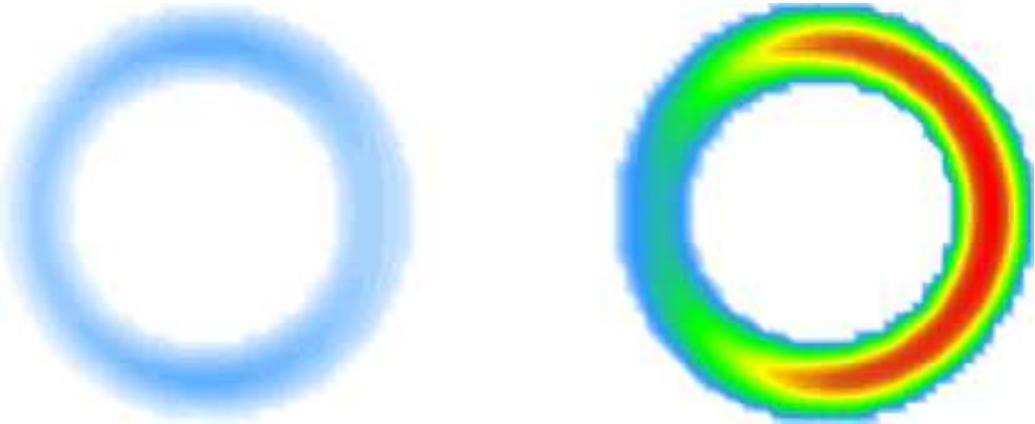} \\
  & & \\
 \includegraphics[width=2.75cm]{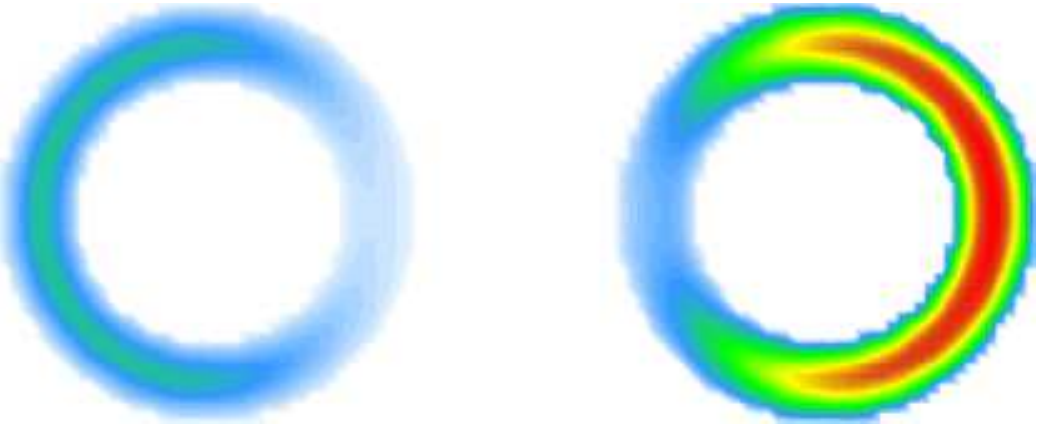} &
 \includegraphics[width=2.75cm]{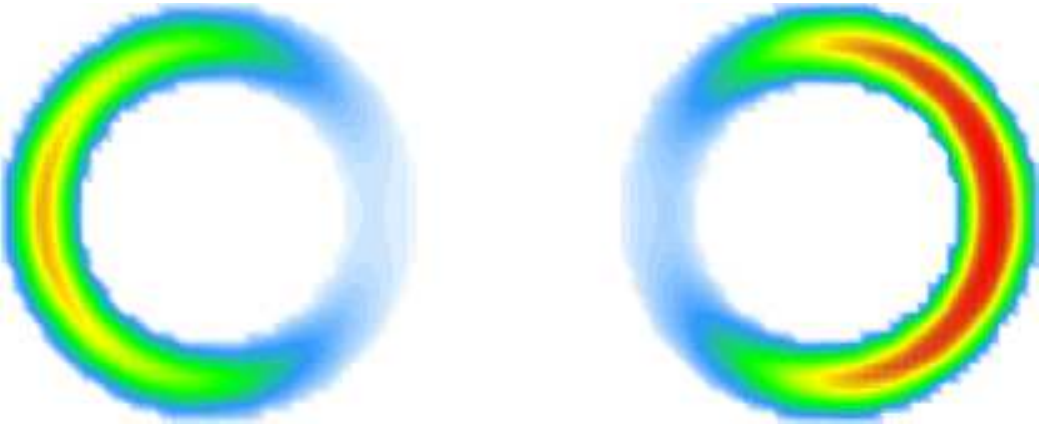} &
 \includegraphics[width=2.75cm]{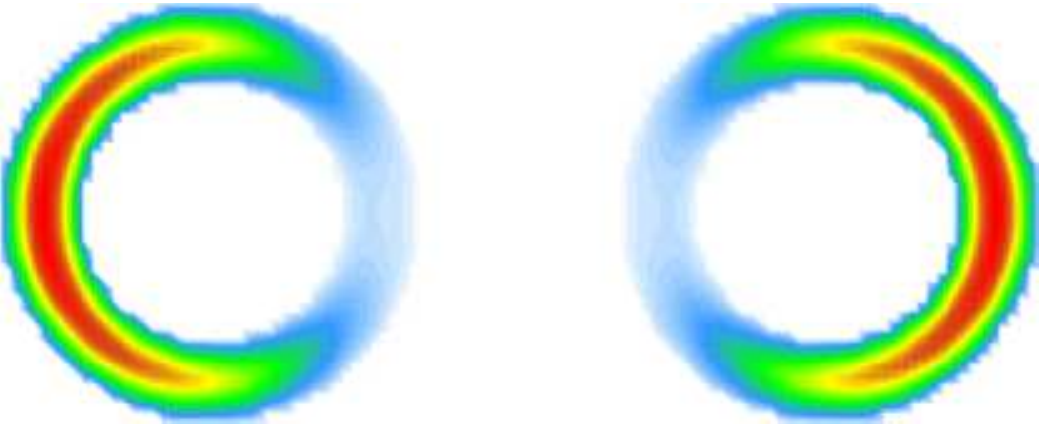} 
\end{tabular} 
   \caption{Concentration of diffusive proteins with curvature preference on the torus using $128\times128\times128$ mesh.  The diffusive 
proteins prefer the curvature $C_0^{\rm pro} = 0.5$ (located on the outer ring) and the background lipid species prefer the 
curvature $C_0^{\rm lip} = -0.1$ (located on the inner ring).  Times shown are $t = 0, 0.1, 0.25, 0.5, 1.0$, and $5.0$.  
The color is scaled by the maximum concentration in each plot.}
   \label{fig:torus_curv128} 
\end{figure}

To drive the diffusive species to the outer ring and the background to the inner ring, we set the spontaneous 
curvatures $C_0^{\rm pro} = 0.5$ and $C_0^{\rm lip} = -0.1$.  The numerical solution to the curvature-driven diffusion 
equation with these spontaneous curvatures is presented in Figure \ref{fig:torus_curv128}.
Recall that under pure diffusion, the diffusive concentration tended toward the inner ring due to numerical error (see Figure \ref{fig:torus2_N128}).  
The curvature effect has this numerical error working against it.  In spite of this, the curvature preference of the diffusive proteins toward 
the outer regions can be clearly seen in Figure \ref{fig:torus_curv128}.

\begin{figure}[!ht]
   \centering
 \begin{tabular}{lll}
 \includegraphics[width=3.125cm]{Torus_N128_t0B}   &  
 \includegraphics[width=3.125cm]{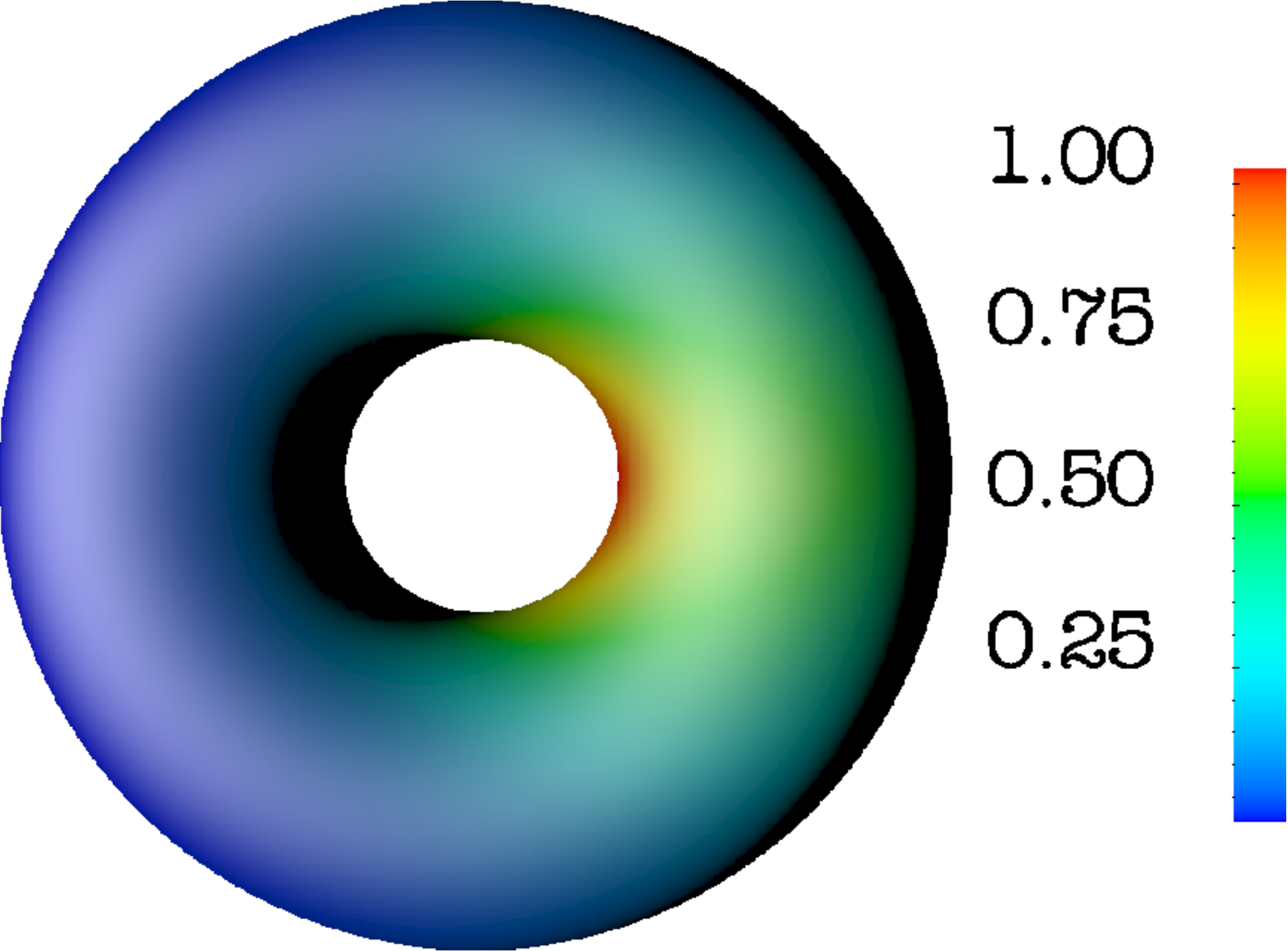} &
 \includegraphics[width=3.125cm]{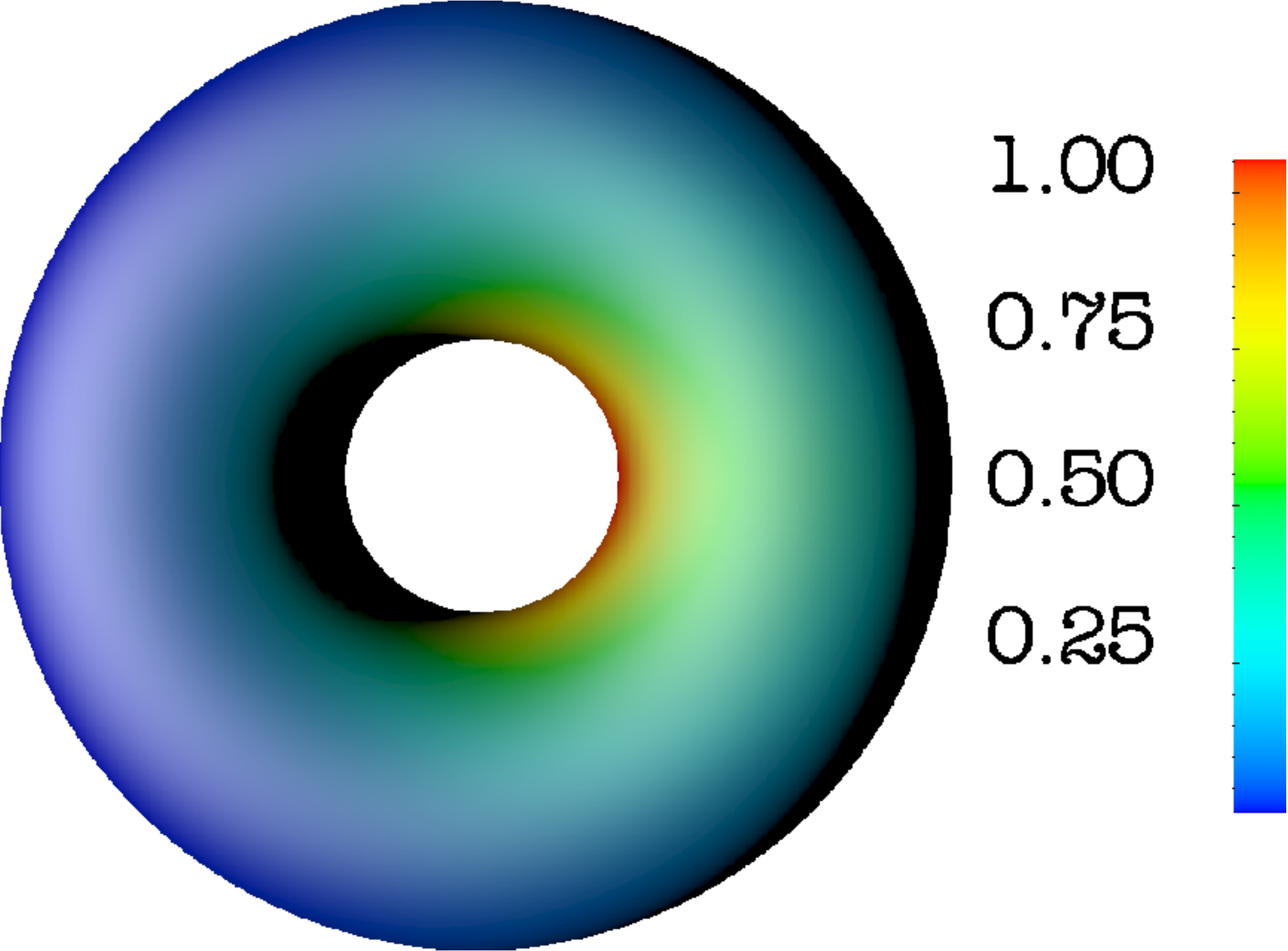} \\ 
 \includegraphics[width=3.125cm]{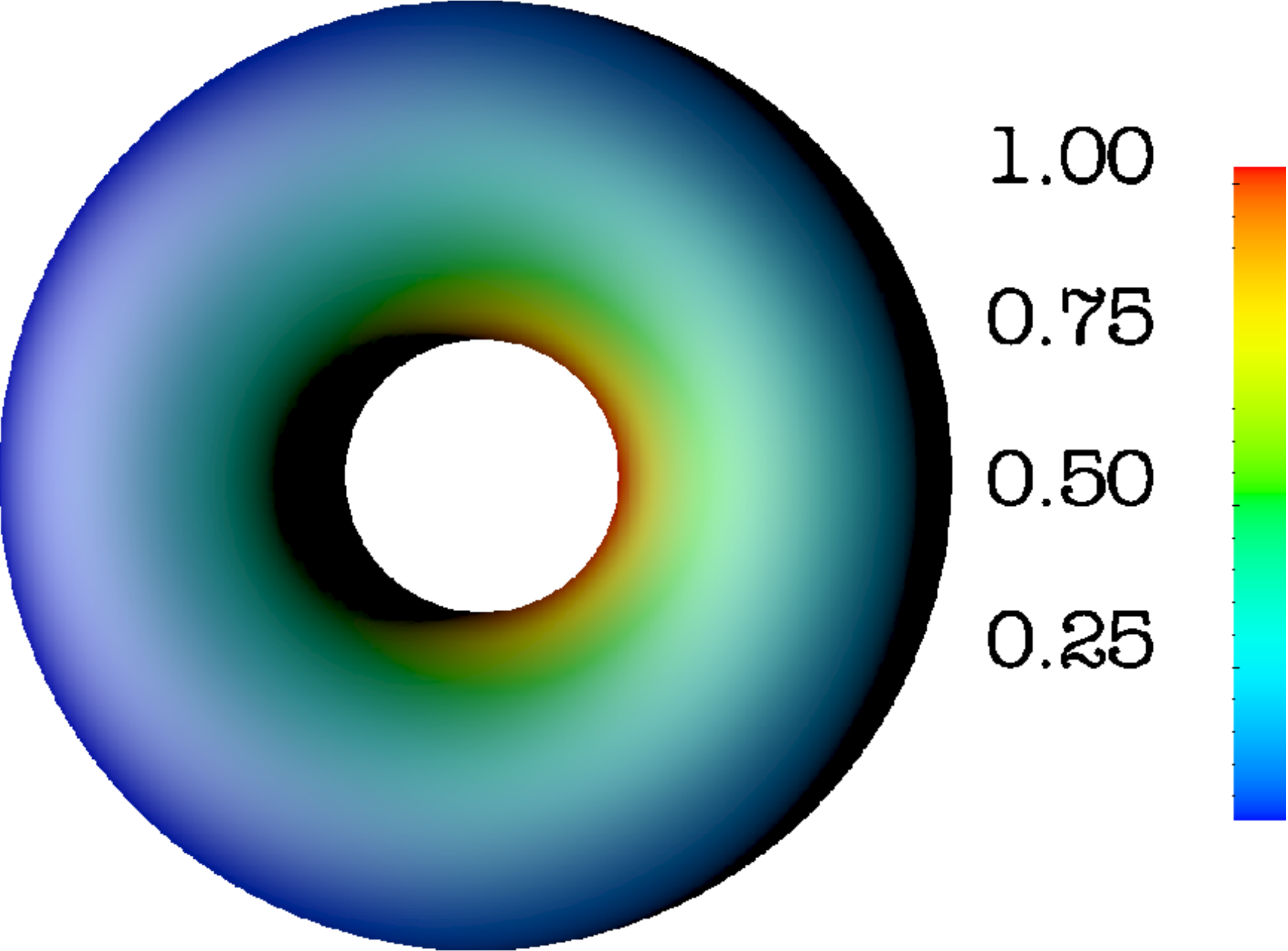} &
 \includegraphics[width=3.125cm]{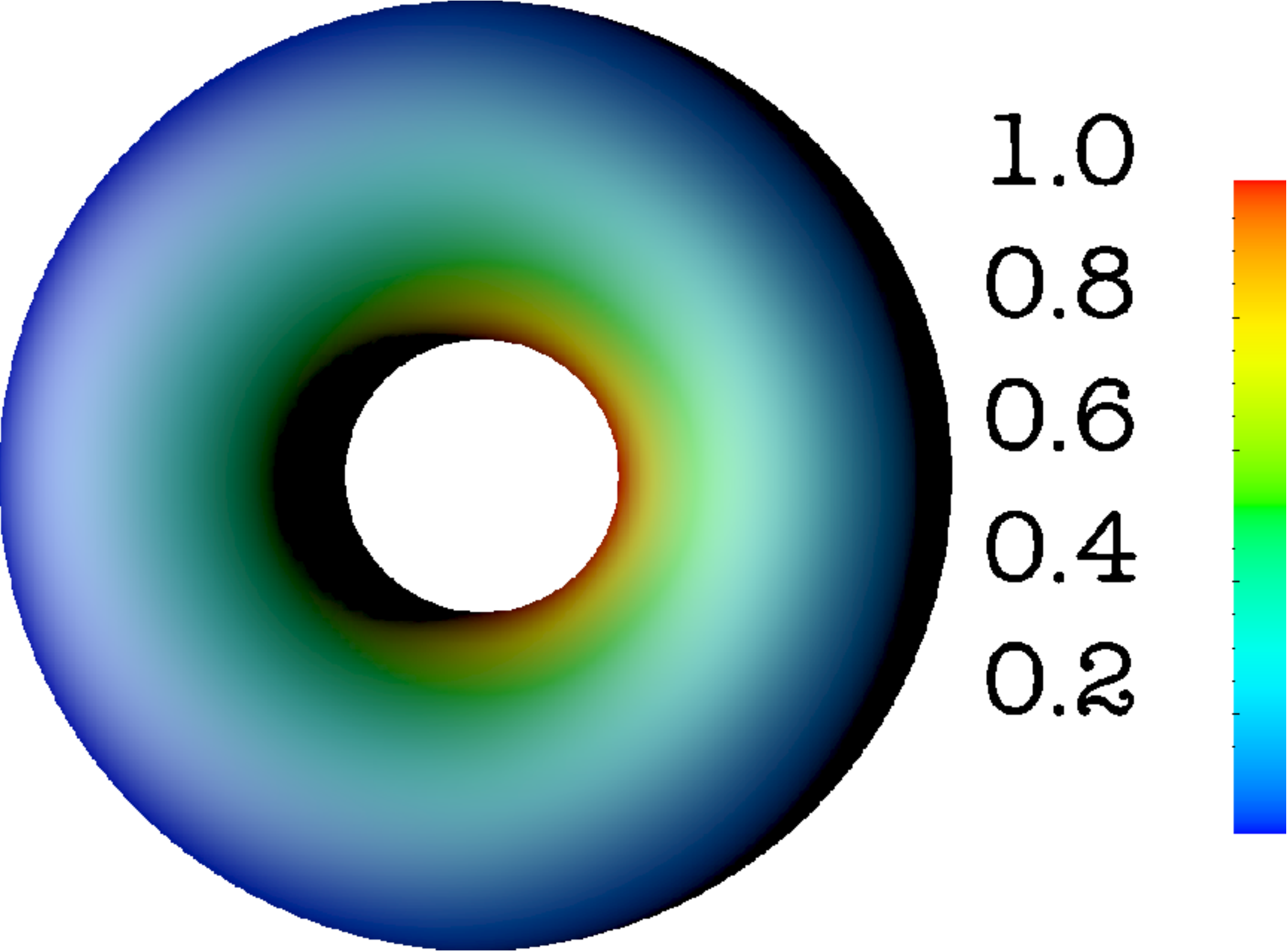} &
 \includegraphics[width=3.125cm]{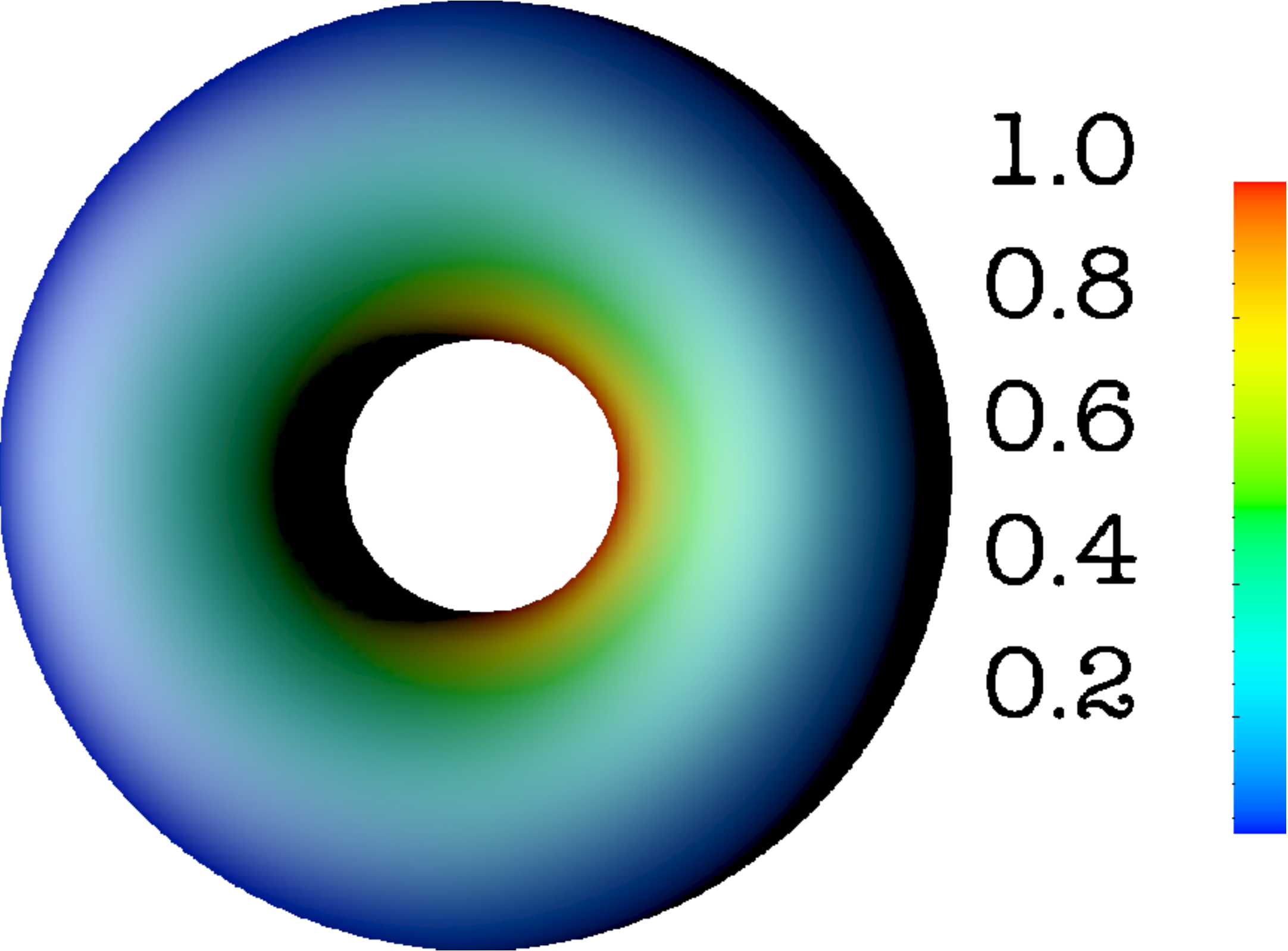} \\
  & & \\
 \includegraphics[width=2.75cm]{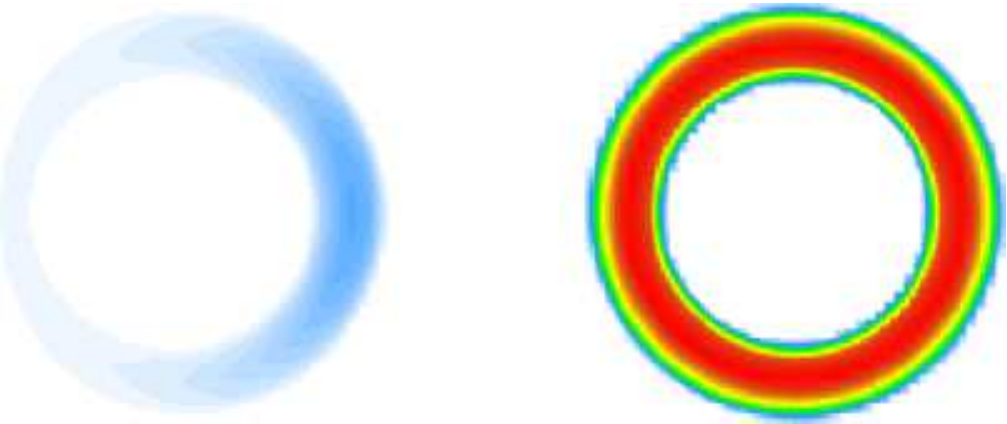}   &  
 \includegraphics[width=2.75cm]{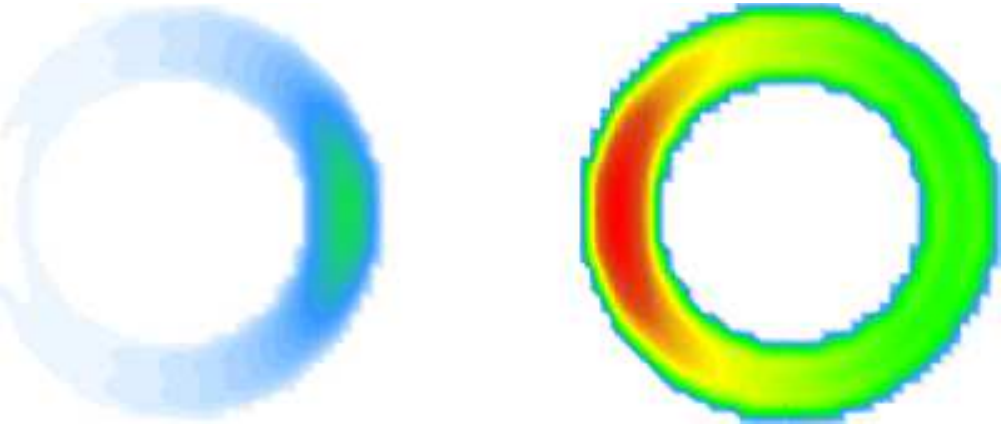} &
 \includegraphics[width=2.75cm]{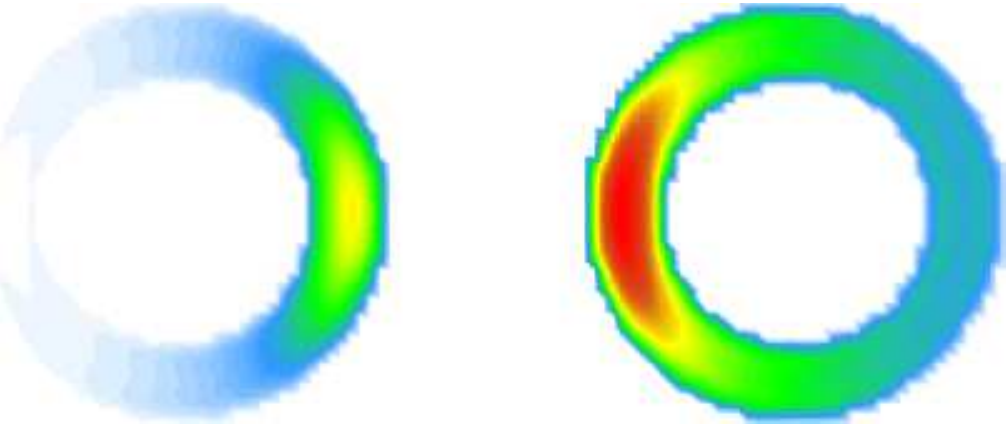} \\ 
  & & \\
 \includegraphics[width=2.75cm]{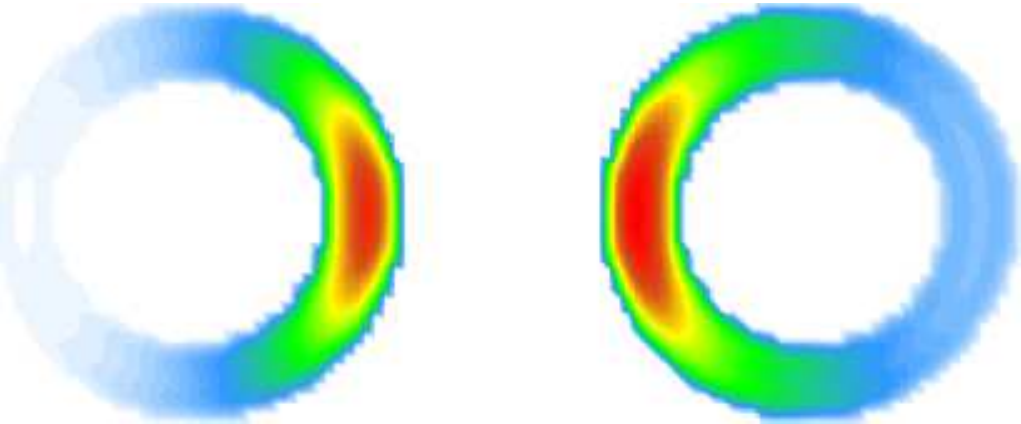} &
 \includegraphics[width=2.75cm]{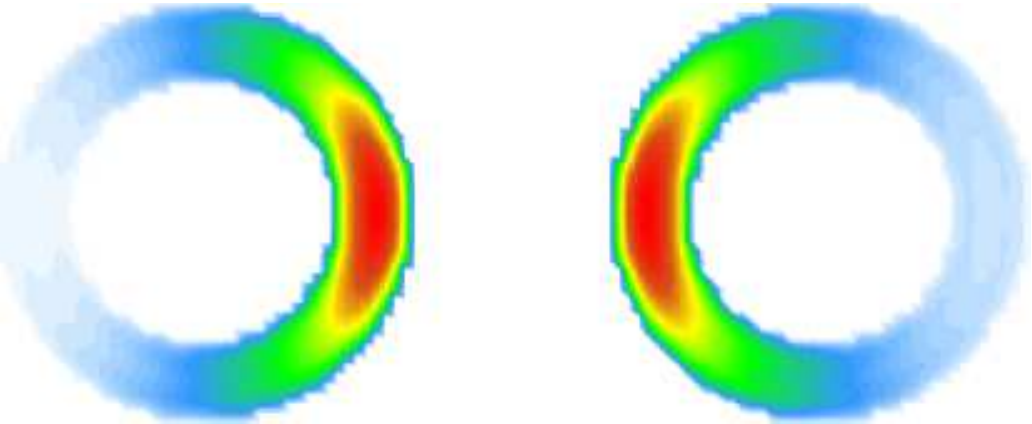} &
 \includegraphics[width=2.75cm]{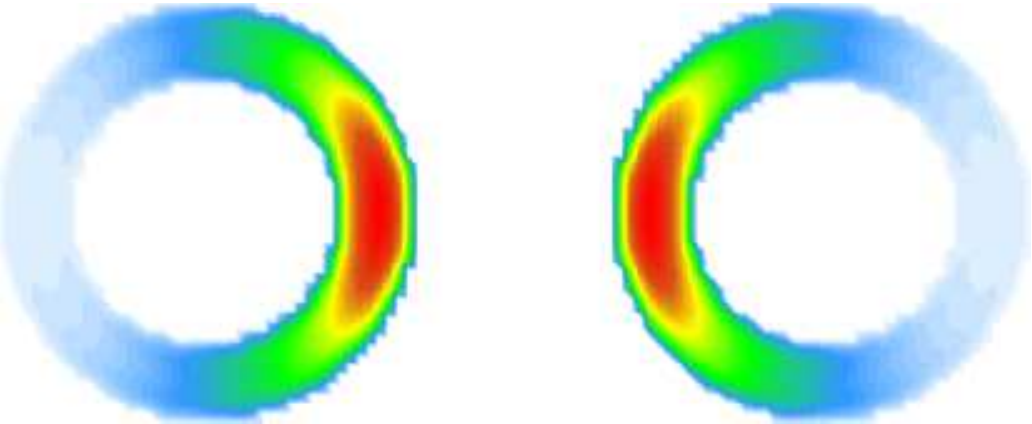} 
 \end{tabular}
   \caption{Concentration of diffusive proteins with curvature preference on the torus using $128\times128\times128$ mesh.  The diffusive proteins 
prefer the curvature $C_0^{\rm pro} = -0.1$ (located on the inner ring) and the background lipid species prefer the curvature $C_0^{\rm lip} = 0.5$ 
(located on the outer ring).  Times shown are $t = 0, 0.1, 0.25, 0.5, 1.0$, and $5.0$.  The color is scaled by the maximum concentration in each plot.}
   \label{fig:torus_curv2_128} 
\end{figure}

Reversing the curvature preference, we now solve the equation with the diffusive proteins $C_0^{\rm pro} = -0.1$ and the 
background lipids $C_0^{\rm lip} = 0.5$.  This is more like the motivating application of the M2 protein, preferring regions of negative 
curvature.  The plots of the concentration with these curvature preferences together with the corresponding cross sections are shown in 
Figure \ref{fig:torus_curv2_128}.  The curvature preference toward the inner ring of the torus can be seen to a much greater 
extent than the numerical error in the pure diffusion case.  
%




\section{Conclusions and future work}\label{sec:conc}
In this paper, we have modeled the curvature preference of diffusion molecules in bilayer membranes using an energetic variational principle.
We introduced an molecular concentration dependent spontaneous curvature in defining the bending energy of the bilayer membrane. This bending
energy and the entropic energy of molecules distribution consist the total energy of the interacted protein-membrane system. The transportation
of molecules on membrane surfaces follow the gradient flow of this total energy. We derived a drift-diffusion equation for the gradient flow,
where the difference between the local membrane curvature and the molecular concentration dependent local spontaneous curvature appears as a 
drift potential. This feature indicates the unification of the biophysics of molecular localization on membrane surfaces and its model developed
in this work. This localization is further demonstrated by the numerical solutions of the drift-diffusion equation on torus. These solutions 
simulate the localization toward the inner and outer rings, corresponding to the different specific intrinsic curvature of the diffusive
molecules.

Various extensions to the curvature-driven model presented can be made. First, the model can be coupled to the evolution of the phase field 
function so the localization on moving membrane surfaces can be simulated. This can be achieved by coupling to gradient flow of the total 
energy to evolve the phase field function \cite{Du2006}. Second, we presented numerical results for a single diffusive species, but the 
model allows for multiple diffusive species. The growth of the computational cost is linear with each additional species, since each 
additional diffusive species requires solving one additional diffusion equation. In addition, we can also consider spatially variable diffusion 
coefficients and the finite size effects ($R$) in our numerical results \cite{Zhou2012}. By including these parameters the simulations can be made 
more realistic and results can be predictive in quantitatively accessing the capability of specific proteins and their mutations in modulating 
membrane curvatures.


\section*{Acknowledgments}
The authors thank Michael Grabe for many helpful discussions.

\bibliographystyle{siam}


\begin{thebibliography}{10}

\bibitem{Adalsteinsson2003}
{\sc D.~Adalsteinsson and J.~A. Sethian}, {\em Transport and diffusion of
  material quantities on propagating interfaces via level set methods}, Journal
  of Computational Physics, 185 (2003), pp.~271--288.

\bibitem{AlmeidaP1995a}
{\sc P.~F.~F. Almeida and W.~L.~C. Vaz}, {\em Lateral diffusion in membranes},
  in Handbook of Biological Physics, Elservier, 1995, ch.~6, pp.~305--.

\bibitem{Antonny2011}
{\sc B.~Antonny}, {\em Mechanisms of membrane curvature sensing}, Annual Review
  of Biochemistry, 80 (2011), pp.~101--123.
\newblock PMID: 21438688.

\bibitem{Baumgart2011}
{\sc T.~Baumgart, B.~R. Capraro, C.~Zhu, and S.~L. Das}, {\em Thermodynamics
  and mechanics of membrane curvature generation and sensing by proteins and
  lipids}, Annual review of physical chemistry, 62 (2011), p.~483.

\bibitem{Baumgart2003}
{\sc T.~Baumgart, S.~T. Hess, and W.~W. Webb}, {\em Imaging coexisting fluid
  domains in biomembrane models coupling curvature and line tension}, Nature,
  425 (2003), pp.~821--824.

\bibitem{Callenberg2012}
{\sc K.~M. Callenberg, N.~R. Latorraca, and M.~Grabe}, {\em Membrane bending is
  critical for the stability of voltage sensor segments in the membrane},
  Journal of General Physiology, 140 (2012), pp.~55--68.

\bibitem{Canham1970}
{\sc P.~Canham}, {\em The minimum energy of bending as a possible explanation
  of the biconcave shape of the human red blood cell}, Journal of Theoretical
  Biology, 26 (1970), pp.~61 -- 81.

\bibitem{Cui2011}
{\sc H.~Cui, E.~Lyman, and G.~A. Voth}, {\em Mechanism of membrane curvature
  sensing by amphipathic helix containing proteins}, Biophysical Journal, 100
  (2011), pp.~1271 -- 1279.

\bibitem{Deckelnick2010}
{\sc K.~Deckelnick, G.~Dziuk, C.~M. Elliott, and C.-J. Heine}, {\em An h-narrow
  band finite-element method for elliptic equations on implicit surfaces},
  {IMA} {J}ournal of {N}umerical {A}nalysis, 30 (2010), pp.~351--376.

\bibitem{Du2005}
{\sc Q.~Du, C.~Liu, R.~Ryham, and X.~Wang}, {\em Modeling the spontaneous
  curvature effects in static cell membrane deformations by a phase field
  formulation}, Communications on Pure and Applied Analysis, 4 (2005),
  pp.~537--548.

\bibitem{Du2004}
{\sc Q.~Du, C.~Liu, and X.~Wang}, {\em A phase field approach in the numerical
  study of the elastic bending energy for vesicle membranes}, Journal of
  Computational Physics, 198 (2004), pp.~450 -- 468.

\bibitem{Du2006}
\leavevmode\vrule height 2pt depth -1.6pt width 23pt, {\em Simulating the
  deformation of vesicle membranes under elastic bending energy in three
  dimensions}, Journal of Computational Physics, 212 (2006), pp.~757 -- 777.

\bibitem{Dziuk1988}
{\sc G.~Dziuk}, {\em Finite elements for the beltrami operator on arbitrary
  surfaces}, in Partial Differential Equations and Calculus of Variations,
  S.~Hildebrandt and R.~Leis, eds., vol.~1357 of Lecture Notes in Mathematics,
  Springer Berlin Heidelberg, 1988, pp.~142--155.

\bibitem{Dziuk2010}
{\sc G.~Dziuk and C.~M. Elliott}, {\em An eulerian approach to transport and
  diffusion on evolving implicit surfaces}, Computing and {V}isualization in
  {S}cience, 13 (2010), pp.~17--28.

\bibitem{Evans1974}
{\sc E.~Evans}, {\em Bending resistance and chemically induced moments in
  membrane bilayers}, Biophysical Journal, 14 (1974), pp.~923--931.

\bibitem{Faraudo2002}
{\sc J.~Faraudo}, {\em Diffusion equation on curved surfaces. {I}. theory and
  application to biological membranes}, Journal of Chemical Physics, 116
  (2002), p.~5831.

\bibitem{Farsad2003}
{\sc K.~Farsad and P.~D. Camilli}, {\em Mechanisms of membrane deformation},
  Curr.\ Opin.\ Cell \ Biol., 15 (2003), pp.~372 -- 381.

\bibitem{Frank1958}
{\sc F.~C. Frank}, {\em I. liquid crystals. on the theory of liquid crystals},
  Discussions of the Faraday Society, 25 (1958), pp.~19--28.

\bibitem{Greer2006}
{\sc J.~B. Greer, A.~L. Bertozzi, and G.~Sapiro}, {\em Fourth order partial
  differential equations on general geometries}, Journal of Computational
  Physics, 216 (2006), pp.~216--246.

\bibitem{Helfrich1973}
{\sc W.~Helfrich et~al.}, {\em Elastic properties of lipid bilayers: theory and
  possible experiments}, Zeitschrift f{\"u}r Naturforschung C, 28 (1973),
  pp.~693--703.

\bibitem{Huttner2001}
{\sc W.~B. Huttner and J.~Zimmerberg}, {\em Implications of lipid microdomains
  for membrane curvature, budding and fission: Commentary}, Current Opinion in
  Cell Biology, 13 (2001), pp.~478--484.

\bibitem{Kim1998}
{\sc K.~S. Kim, J.~Neu, and G.~Oster}, {\em Curvature-mediated interactions
  between membrane proteins}, Biophysical Journal, 75 (1998), pp.~2274--2291.

\bibitem{Lee2012}
{\sc H.~G. Lee and J.~Kim}, {\em Regularized {D}irac delta functions for phase
  field models}, International Journal for Numerical Methods in Engineering, 91
  (2012), pp.~269--288.

\bibitem{Olshanskii2015}
{\sc M.~A. Olshanskii and D.~Safin}, {\em A narrow-band unfitted finite element
  method for elliptic pdes posed on surfaces}.
\newblock Preprint.

\bibitem{Osher2003}
{\sc S.~Osher and R.~Fedkiw}, {\em Level Set Methods and Dynamic Implicit
  Surfaces}, vol.~153 of Applied Mathematical Sciences, Springer, 2003.

\bibitem{Parton2011}
{\sc D.~L. Parton, J.~W. Klingelhoefer, and M.~S. Sansom}, {\em Aggregation of
  model membrane proteins, modulated by hydrophobic mismatch, membrane
  curvature, and protein class}, Biophysical Journal, 101 (2011), pp.~691 --
  699.

\bibitem{Rossman2010}
{\sc J.~S. Rossman, X.~Jing, G.~P. Leser, V.~Balannik, L.~H. Pinto, and R.~A.
  Lamb}, {\em Influenza virus {M2} ion channel protein is necessary for
  filamentous virion formation}, Journal of Virology, 84 (2010),
  pp.~5078--5088.

\bibitem{Schmidt2013}
{\sc N.~W. Schmidt, A.~Mishra, J.~Wang, W.~F. DeGrado, and G.~C.~L. Wong}, {\em
  Influenza virus {A} {M2} protein generates negative gaussian membrane
  curvature necessary for budding and scission}, Journal of the American
  Chemical Society, 135 (2013), pp.~13710--13719.

\bibitem{Slotboom1973}
{\sc J.~W. Slotboom}, {\em Computer-aided two-dimensional analysis of bipolar
  transistors}, Electron Devices, IEEE Transactions on, 20 (1973),
  pp.~669--679.

\bibitem{SodtA2014a}
{\sc A.~J. Sodt and P.~R. W.}, {\em Molecular modeling of lipid membrane
  curvature induction by a peptide: More than simply shape}, Biophysical
  Journal, 106 (2014), pp.~1958--1969.

\bibitem{Stachowiak2010}
{\sc J.~C. Stachowiak, C.~C. Hayden, and D.~Y. Sasaki}, {\em Steric confinement
  of proteins on lipid membranes can drive curvature and tubulation},
  Proceedings of the National Academy of Sciences, 107 (2010), pp.~7781--7786.

\bibitem{Strain1994}
{\sc J.~Strain}, {\em Fast spectrally-accurate solution of variable-coefficient
  elliptic problems}, Proceedings of the American Mathematical Society, 122
  (1994), pp.~843--850.

\bibitem{Teigen2011}
{\sc K.~E. Teigen, P.~Song, J.~Lowengrub, and A.~Voigt}, {\em A
  diffuse-interface method for two-phase flows with soluble surfactants},
  Journal of Computational Physics, 230 (2011), pp.~375 -- 393.

\bibitem{Du2008}
{\sc X.~Wang and Q.~Du}, {\em Modelling and simulations of multi-component
  lipid membranes and open membranes via diffuse interface approaches}, Journal
  of Mathematical Biology, 56 (2008), pp.~347--371.

\bibitem{Zhao2003}
{\sc J.-J. Xu and H.-K. Zhao}, {\em An {E}ulerian formulation for solving
  partial differential equations along a moving interface}, Journal of
  Scientific Computing, 19 (2003), pp.~573--594.

\bibitem{Zhou2012}
{\sc Y.~C. Zhou}, {\em Electrodiffusion of lipids on membrane surfaces},
  Journal of Chemical Physics, 136 (2012), p.~205103.

\bibitem{Zimmerberg2006}
{\sc J.~Zimmerberg and M.~M. Kozlov}, {\em How proteins produce cellular
  membrane curvature}, Nature Reviews Molecular Cell Biology, 7 (2006),
  pp.~9--19.

\end{thebibliography}
 
\end{document}